\documentclass[12pt,preprint]{aastex}

\usepackage{graphicx}
\usepackage{multirow}

\newcommand{\asca}{{\small \it ASCA}}

\newcommand{\rosat}{{\small \it ROSAT}}

\newcommand{\agn}{{\small AGN}}
\newcommand{\ngc}{{\small NGC~3783}}
\newcommand{\mcg}{{\small MCG~--6--30--15}}

\newcommand{\hullac}{{\small HULLAC}}
\newcommand{\cloudy}{{\small CLOUDY}}
\newcommand{\xstar}{{\small XSTAR}}
\newcommand{\fwhm}{{\small FWHM}}

\newcommand{\xmm}{{\it XMM-Newton}}
\newcommand{\rgs}{{\small RGS}}
\newcommand{\chandra}{{\it Chandra}}
\newcommand{\suzaku}{{\it Suzaku}}

\newcommand{\hetgs}{{\small HETGS}}

\newcommand{\uv}{{\small UV}}
\newcommand{\x}{X-ray}

\newcommand{\cmsq}{cm$^{-2}$}
\newcommand{\kms}{km~s$^{-1}$}
\newcommand{\iras}{IRAS 13349+2438}
\newcommand{\cgs}{erg~s$^{-1}$~cm}

\begin{document}

\title{X-Ray Absorption Analysis of MCG~--6-30-15: \\
Discerning Three Kinematic Systems}
\author{Tomer Holczer \altaffilmark{1}, Ehud Behar \altaffilmark{2,3}, and
Nahum Arav \altaffilmark{4}}

\altaffiltext{1}{Department of Physics,
                 Technion, Haifa 32000, Israel.
                 tomer@physics.technion.ac.il (TH),
                 behar@physics.technion.ac.il (EB).}
\altaffiltext{2}{ Code 662, NASA / Goddard Space Flight Center,
Greenbelt, MD 20771, USA} \altaffiltext{3}{ Senior NPP Fellow on
leave from the Technion, Israel} \altaffiltext{4}{Department of
Physics, Virginia Tech, Blacksburg, Va 24061. arav@vt.edu }

\received{} \revised{} \accepted{}

\shorttitle{AMD of \iras } \shortauthors{Holczer et al.}

\begin{abstract}
By analyzing the \x\ spectrum of \mcg\ obtained with the \hetgs\
spectrometer on board the \chandra\ observatory, we identify three
kinematically distinct absorption systems; two outflow components
intrinsic to \mcg, and one local at $z = 0$. The slow outflow at
$-100~\pm\ 50$~\kms\ has a large range of ionization manifested by
absorption from 24 different charge states of Fe, which enables a
detailed reconstruction of the absorption measure distribution
($AMD$). This $AMD$ spans five orders of magnitude in ionization
parameter: $-1.5 < \log\xi < 3.5$ (\cgs), with a total column
density of $N_H$ = (5.3~$\pm$ 0.7) $\times10^{21}$~\cmsq. The fast
outflow at $-1900~\pm 150$~\kms\ has a well defined ionization
parameter with $\log \xi = 3.82\pm 0.03$ (\cgs) and column density
$N_H~= 8.1~\pm\ 0.7 \times10^{22}$~\cmsq. Assuming this component
is a thin, uniform, spherical shell, it can be estimated to lie
within 11 light days of the \agn\ center. The third component,
most clearly detected in the lower oxygen charge states O$^{+1}$ -
O$^{+6}$, has been confused in the past with the fast outflow, but
is identified here with local gas ($z = 0$) and a total column
density $N_H$ of a few $10^{20}$~\cmsq. Finally, we exploit the
excellent spectral resolution of the \hetgs\ and use the present
spectrum to determine the rest-frame wavelengths  of oxygen
inner-shell lines that were previously uncertain.

\end{abstract}


\clearpage
\section{INTRODUCTION}
\label{sec:intro}

Approximately half of type 1 \agn s show complex absorption in the
soft \x\ band \citep{crenshaw03}. The blue shifted absorption
lines come from a highly or partly ionized outflow first noted by
\citet[]{halpern84}. These outflows may play a central role in
cosmological feedback, in the metal enrichment of the
intergalactic medium (IGM) and in understanding black hole
evolution. However, the outflow physical properties such as mass,
energy and momentum are still largely unknown.A necessary step to
advance on these issues is to obtain reliable measurements of the
ionization distribution, and column densities of the outflowing
material.

\mcg\ is a bright Seyfert~1 galaxy ($L_X\sim10^{43}$ erg~s$^{-1}$)
at a redshift z~=~0.007749 \citep{fisher95}. It shows strong
optical reddening \citep{reynolds97}, which suggests that the
absorber may also contain some cold (neutral) gas. \mcg~ is a
highly variable \x\ source with changes up to a factor of 2 on
time scales of $\sim$ 1 ks \citep{fabian94,reynolds95,otani96}
\mcg\ is well known for its relativistically broadened emission
feature at 5~-- 7 keV \citep[]{tanaka95,ba03,young05}  that has
been interpreted as Fe K$\alpha$ emission from the accretion disk
deep inside the black hole gravitational potential well.

There are numerous studies in the literature of the ionized \x\
absorber of \mcg\ the most important of which are listed in
Table~\ref{table1}. \citet{otani96} analyzed the \asca\ data and
\citet{reynolds97} fitted the \rosat\ and \asca\ spectra using
strong, variable, oxygen absorption edges to explain the curved
continuum shape. In 2000, \citet{brand01} obtained the first
grating spectra of \mcg\ with the Reflection Grating Spectrometer
(\rgs) on board \xmm. The improved resolution showed the spectral
turn overs to be inconsistent with the oxygen edge positions.
Indeed, \citet{brand01} modeled the soft \x\ spectrum with
relativistic emission lines from deep in the accretion disk of
C$^{+5}$, N$^{+6}$ and O$^{+7}$, instead of absorption edges.
\citet{lee01} then used the 2000 \chandra~ High Energy
Transmission Grating (\hetgs) spectrum of \mcg\ to claim that
(O$^{+6}$ but mostly) neutral Fe L-shell edges can explain the
continuum turn over at $\approx 17.5$~\AA\ \citep[see
also][]{lee09}, and not relativistic emission lines. As in
previous works, \citet{lee01} needed two ionization zones in order
to fit the soft \x\ absorption lines. \citet{sako03} reinstated
the \citet{brand01} interpretation of the 2000 \rgs\ data, but
with a better absorption model. \citet{sako03} identified two
velocity components in the absorber with outflow velocities of
--150 \kms\ and --1900 \kms . \citet{turner04} fitted a second
longer 320~ks \rgs\ observation from 2001, and confirmed the two
kinematic components, with (roughly consistent) velocities of
+80~$\pm$ 260~\kms\ and --1970~$\pm$ 160~\kms. \citet{young05}
studied a long 520 ks 2004 \chandra\ \hetgs\ observation, and
found some of the most highly ionized species of Fe$^{+24}$,
Fe$^{+25}$, S$^{+15}$ and Si$^{+13}$ to be outflowing at --2000
\kms\ . The analysis of \citet{mckernan07} of the previous 2000
\hetgs\ spectrum again confirmed the fast component, but with a
slightly different velocity of $-1550^{+80}_{-130}$ \kms. They
also reported two slow components with low velocities of
$\leq$~30~\kms\ and $\leq$~15 \kms. Recently, \citet{miller08}
studied all of the archival spectra of \mcg\ including a 2006
\suzaku\ observation and confirmed the fast outflow and the slow
outflow with its two ionization components. \citet{miller08} also
invoked a partial covering absorber that helps explain the
continuum shape without the need of relativistically broadened
emission lines. Most recently, \citet{chelouche08} modeled the
2004 \hetgs\ spectrum with similar components as the previous
authors. We summarize the long list of previous works on this
intriguing target and in particular its grating observations by
stating that while the gratings have led to unambiguous
measurements of the kinematics and ionization of the absorbing
outflow, the physical interpretation of the \x\ continuum is still
being debated.

In this paper, we wish to further investigate the physical
conditions of the \mcg\ \x\ absorber, using the archival 520 ks
\hetgs\ observation from 2004, but with special focus on the
ionization distribution of the plasma, and on lines that do not
seem to fit the simple two-velocity picture \citep[see list of
individual-line velocities in][]{turner04}. While the majority of
works on the \x\ spectra of \agn\ outflows employ gradually
increasing number of ionization components, until the fit is
satisfactory \citep[e.g.,][]{kaspi01, sako03}, it is instructive
to reconstruct the actual distribution of the column density in
the plasma as a continuous function of ionization parameter $\xi$
\citep{katrien05}, which we termed the absorption measure
distribution \citep[$AMD$,][]{tomer07}. Although for high-quality
spectra such as the present one, two or three ionization
components might produce a satisfactory fit, the $AMD$
reconstruction is the only method that reveals the actual
distribution including its physical discontinuities (e.g., due to
thermal instability), and ultimately provides a more precise
measurement of the total column density.

\section{DATA REDUCTION}
\label{sec:data}

\mcg\ was observed by \chandra /\hetgs\ on 19--28 May, 2004 for a
total exposure time of 520~ks. All observations were reduced from
the \chandra\ archive using the standard pipeline software (CIAO
version 3.2.1). The total number of counts in the first (plus and
minus) orders between 2 and 20~\AA\ is 353485 for MEG (medium
energy grating) and 165698 for HEG (high energy). More details on
the observation can be found in Table.~\ref{table2}. No background
subtraction was required as the background level was negligible.
Flux spectra were obtained by first co-adding count spectra from
the different refraction orders and convoluting with the broadest
line spread function (MEG $\pm$ 1 orders) to ensure uniformity.
The total count spectra were then divided by the total effective
area curve (summed over orders) and observation time. Finally,
spectra were corrected for neutral Galactic absorption of $N_H$ =
4.1$\times10^{20}$ \cmsq\ \citep{dickey90}.

\section{SPECTRAL MODEL}
\label{sec:modeling}

\begin{figure}

\centering
\includegraphics[width=13.0cm]{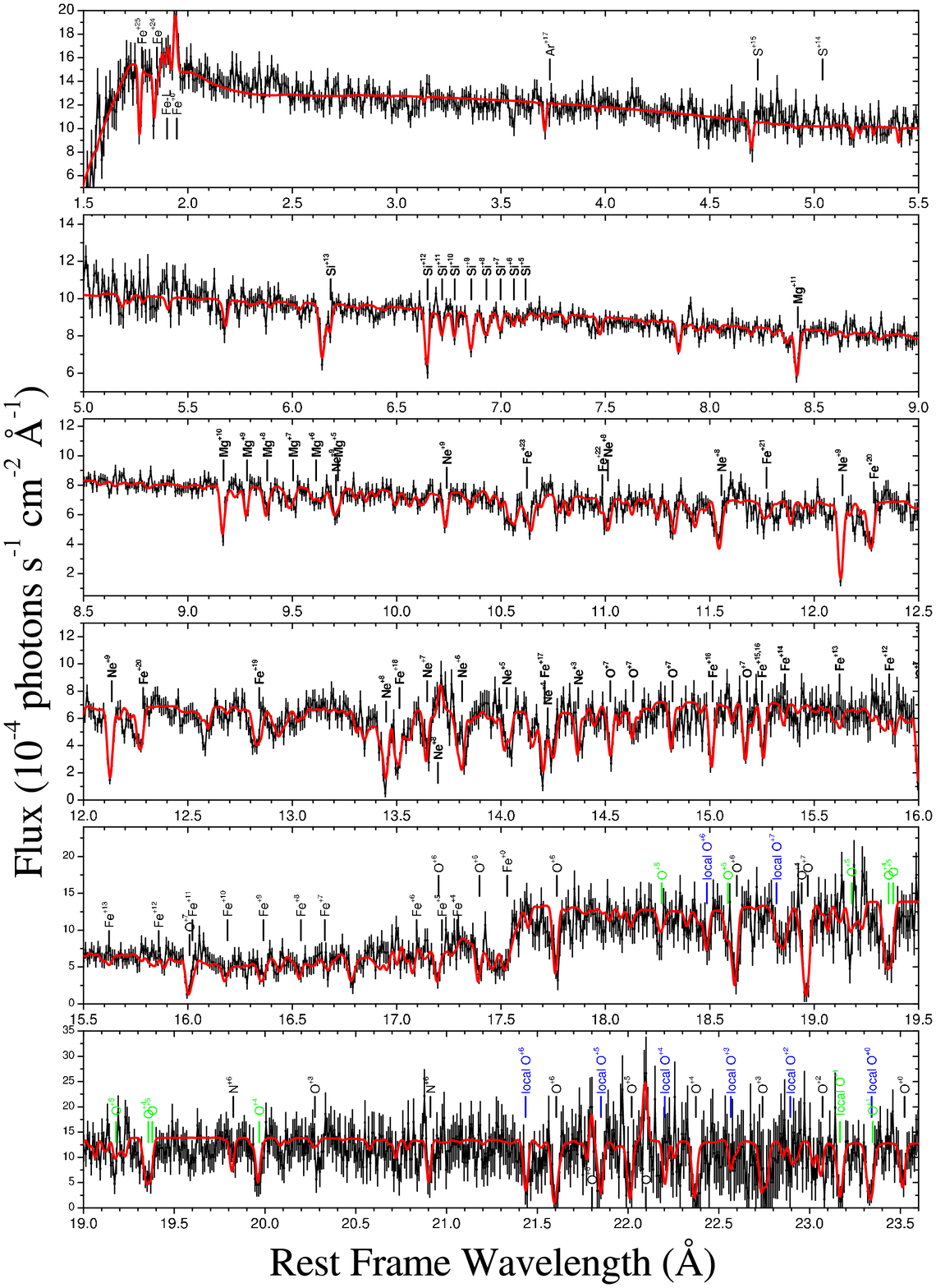}
\caption{\chandra\ \hetgs\ spectrum of \mcg\ corrected for
cosmological redshift ($z$ = 0.007749). Top panel includes only
HEG data, while all other panels present the combined MEG+HEG
spectrum. The red line is the best-fit model including the slow
(\S \ref{sec:amd}) and fast (\S \ref{fast}) outflow components as
well as the local $z = 0$ component (\S \ref{local}). Ions
producing the strongest absorption (emission) lines and blends are
marked above (below) the data. Oxygen lines that require slight
wavelength adjustments are marked in green.} \label{figure1}
\end{figure}

Variations of approximately 40\% on time scales of 20~ks were
observed over the 520~ks total exposure of \mcg. Light curves can
be found in \citet{young05}. The average flux of \mcg\ does not
vary by more than 40$\%$ between maximum and minimum flux levels
over the 14 years it has been observed, as can be seen from the
continuum flux levels quoted in Table~\ref{table1}. The present
work deals with the long term properties of the ionized absorber.
With these small and rapid variations the absorber is not expected
to significantly vary. Henceforth, we generally use the combined
MEG and HEG full 520~ks spectrum that is shown in
Fig.~\ref{figure1}. At the shortest wavelengths ($\lambda <$ 6),
we exploit the superior spectral resolution and effective area of
HEG (top panel in Fig.~\ref{figure1}). The present fitting
procedure follows our ion-by-ion fitting method \citep{behar01,
sako01, behar03, tomer05}. First, we fit for the broad-band
continuum. Subsequently, we fit the absorption features using
template ionic spectra that include all of the absorption lines
and photoelectric edges of each ion, but vary with the broadening
(so-called turbulent) velocity and the ionic column density.
Strong emission lines are fitted as well.

\subsection{Continuum Parameters}
\label{sec:continuum} The continuum \x\ spectrum of most \agn s
can be characterized by a high-energy power-law and a soft excess
that rises above the power law at lower energies below
$\sim$1~keV. This soft excess is often modeled with a blackbody,
or modified blackbody, although it clearly is more spectrally
complex and possibly includes prominent atomic features. \mcg\ can
not be properly fitted by such a simplistic model because of a
sharp jump in flux at $\sim 17.5$~\AA. As explained in \S
\ref{sec:intro}, there is a controversy regarding how to interpret
this sharp spectral feature; One model \citep{brand01,sako03} uses
relativistically broadened emission lines, while the other option
is to invoke a steep soft excess \citep{lee01} that is then
absorbed by a large column density. In this work, given the
ambiguity of the 17.5 \AA\ feature, and since we wish to focus on
the absorption lines, we use a phenomenological cubic spline
continuum, not very different from that used by \citet{turner04},
that characterizes the continuum flux level most adequately
throughout the spectrum.

\subsection{The Ionized Absorber}

The intensity spectrum $I_{ij}(\nu )$ around an atomic absorption line $i
\rightarrow j$ can be expressed as:

\begin{equation}
I_{ij}(\nu) = I_0(\nu)~e^{-N_{ion}\sigma_{ij}(\nu)}
\end{equation}

\noindent where $I_0(\nu)$ represents the unabsorbed continuum intensity,
$\sigma_{ij}(\nu)$ denotes the line absorption cross section for
photo-excitation (in cm$^2$) from ground level $i$ to excited level $j$.
If all ions are essentially in the ground level,
$N_{ion}$ is the total ionic column density towards the source (in cm$^{-2}$).
The photo-excitation cross section is given by:

\begin{equation}
\sigma_{ij}(\nu) = \frac{\pi e^2}{m_ec}f_{ij}\phi(\nu)
\end{equation}

\noindent where the first term is a constant that includes the
electron charge $e$, its mass $m_e$, and the speed of light $c$.
The absorption oscillator strength is denoted by $f_{ij}$, and
$\phi(\nu)$ represents the Voigt profile due to the convolution of
natural (Lorentzian) and Doppler (Gaussian) line broadening. The
Doppler broadening consists of thermal and turbulent motion, but
in \agn~outflows, the turbulent broadening is believed to dominate
the temperature broadening. The Natural broadening becomes
important when the lines saturate as in our current spectrum, e.g.
the O$^{+6}$ line. Transition wavelengths, natural widths and
oscillator strengths were calculated using the Hebrew University
Lawrence Livermore Atomic Code \citep[\hullac,][]{bs01}.
Particularly important for \agn\ outflows are the inner-shell
absorption lines \citep{uta01, behar02}. More recent and improved
atomic data for the Fe M-shell ions were incorporated from
\citet{gu06}.

Since the absorbing gas is outflowing, the absorption lines are
slightly blue-shifted with respect to the \agn\ rest frame.
Although blue shifts of individual lines can differ to a small
degree, we can identify two overall distinct kinematic components
with best-fit outflow velocities of $v = -100~\pm 50$~\kms\ and $v
= -1900~\pm 150$~\kms. These velocities are set in the model to
one value (for each component) for all of the ions. There are
absorption lines mostly from oxygen in the range above $\sim
20$~\AA, which appear even more blue shifted than $-1900$~\kms.
These lines are addressed in detail in \S \ref{local}.

For the slow component, a turbulent velocity of $v_\mathrm{turb}$
= 100~\kms\ is used. The turbulent velocity (referred to by some
as the $b$ parameter) is defined as $v_\mathrm{turb}
$~=~$\sqrt{2}\sigma$~=~FWHM/$\sqrt{4\ln2}$, where $\sigma$ is the
standard deviation and FWHM is the full width at half max. The
value of 100~\kms , although, unresolved by \hetgs , provides a
good fit to the strongest absorption lines in the spectrum. For
the fast component, $v_\mathrm{turb}$~= 500~\kms\ (\fwhm ~=
830~\kms) is used. Since the fast component is detected mostly in
highly-ionized species at short wavelengths where the grating
resolving power is lowest, this value is obtained primarily from
the O$^{+7}$ Ly\,$\alpha$ line at 18.97~\AA, but fits all lines
well. Finally, the model includes also the 23~m\AA\ instrumental
broadening, which in terms of velocity (\fwhm) is, e.g., 460~\kms\
for the Fe$^{+16}$ resonant line at 15.01~\AA\ and 1115~\kms\ for
the Si$^{+13}$ Ly\,$\alpha$ line at 6.18~\AA. The value of
$v_\mathrm{turb}$~= 100~\kms\ for the slow component is the same
value used by \citet{lee01} and by \citet{sako03}.
\citet{mckernan07} use 170~\kms , which is still just unresolved
by the \hetgs . The value of $v_\mathrm{turb}$~= 500 \kms\ for the
fast component is consistent with that quoted by \citet{young05}
and by \citet{miller08}.

Our model includes all of the important lines of all ion species
that can absorb in the waveband observed by \hetgs. In the slow
component of \mcg, we find evidence for the following ions:
N$^{+6}$, Fe$^{+1}$--~Fe$^{+23}$ as well as neutral iron, all
oxygen stats, Ne$^{+3}$--~Ne$^{+9}$, Mg$^{+4}$--~Mg$^{+11}$ and
Si$^{+5}$--~Si$^{+13}$. We also include the K-shell photoelectric
edges for all these ions although their effect here is largely
negligible. When fitting the data, each ionic column density is
treated as a free parameter. A preliminary spectral model is
obtained using a Monte-Carlo fit applied to the entire spectrum.
Subsequently, the final fit is obtained for individual ionic
column densities in a more controlled manner, which ensures that
the fit of the leading lines is not compromised. The best fit
model is shown in Fig.~\ref{figure1}. It can be seen that most
ions are reproduced fairly well by the model. Note that some lines
could be saturated, e.g., the leading lines of O$^{+6}$ and
O$^{+7}$. In these cases, the higher order lines with lower
oscillator strengths are crucial for obtaining reliable $N_{ion}$
values.

\subsection{AMD Method}
\label{sec:$AMD$}

The large range of ionization states present in the absorber
strongly suggests that the absorption arises from gas that is
distributed over a wide range of ionization parameter $\xi$.
Throughout this work, we use the following convention for the
ionization parameter $\xi = L / (n_Hr^2)$ in units of \cgs , where
$L$ is the ionizing luminosity, $n_H$ is the H number density, and
$r$ is the distance from the ionizing source. We apply the
Absorption Measure Distribution ($AMD$) analysis in order to
obtain the total hydrogen column density $N_H$ along the line of
sight. The $AMD$ can be expressed as:

\begin{equation}
AMD \equiv \partial N_H/ \partial(\log \xi)
\end{equation}

and

\begin{equation}
N_H = \int AMD~d(\log \xi)
\end{equation}

\noindent The relation between the ionic column densities
$N_{ion}$ and the $AMD$ is then expressed as:

\begin{equation}
N_{ion} = A_z\int\frac{\partial
N_H}{\partial(\log\xi)}f_{ion}(\log \xi)d(\log\xi)
 \label{eqAMD}
\end{equation}

\noindent where $N_{ion}$ is the measured ion column density,
$A_z$ is the element abundance with respect to hydrogen taken from
\citet{as09} and assumed to be constant throughout the absorber,
and $f_{ion}(\log\xi)$ is the fractional ion abundance with
respect to the total abundance of its element. Here, we aim at
recovering the $AMD$ for \mcg.

For the $AMD$, we need to find a distribution $\partial N_H /
\partial(\mathrm{\log}\xi)$ that after integration
(eq.~\ref{eqAMD}) will produce all of the measured ionic column
densities. When fitting an $AMD$, one must take into account the
full dependence of $f_{ion}$ on $\xi$. We employ the \xstar\ code
\citep{kal01} version 2.1kn3 to calculate $f_{ion}(\log\xi)$ using
a best-fit power-law and blackbody continuum extrapolated to the
range of 1 -- 1000 Rydberg. We assume all charge states see the
same ionizing spectrum. This is justified by the absence of
significant bound-free absorption edges in the spectrum. All
elements are expected to reflect the same $AMD$ distribution, due
to the assumption that they all reside in the same gas. Iron
however, has a special role as it covers almost five orders of
magnitude in $\xi$, more than any other element. More details on
the $AMD$ binning method and error calculations can be found in
\citet{tomer07}. Further physical implications emanating form
$AMD$ analysis of Seyfert outflows can be found in Behar (2009,
submitted).

\subsection{Narrow Emission Lines}

The present \mcg\ spectrum has a few narrow, bright emission lines
of Fe, Ne, and O, which are assumed not to be absorbed by the
outflow, but are absorbed by the local component discussed in \S
\ref{local} (and by the neutral Galactic column). These lines are
fitted with simple Gaussians and are found to be stationary to
within $\approx$~70~\kms. The Fe K$\alpha$ blends appear slightly
broader (FWHM = 15 m\AA) than the Ne and O lines whose widths are
consistent with a kinematic broadening of 235 \kms\ FWHM, as
expected for features comprised of many lines from several charge
states. The centroid wavelength and photon flux are measured for
each feature and listed in Table~\ref{table3}. The K$\alpha$
emission by neutral Fe, or generally M-shell Fe ions, is detected
at 1.94~\AA\ (6.4 keV). Weaker K$\alpha$ emission from more highly
ionized L-shell Fe is detected at somewhat shorter wavelengths.
Both of these are likely due to a moderately ionized medium
excited by the continuum. The Ne$^{+8}$ K$\alpha$ forbidden line
at 13.7~\AA\ is less prominent, but can still be detected. The
O$^{+6}$ K$\alpha$ forbidden and intercombination lines at
22.1~\AA, and at 21.8~\AA, respectively, are clearly detected.
Conversely, the He-like {\it resonance} K$\alpha$ lines of Ne and
O are not observed in emission. We believe these emission lines do
exist since they are implied by the other He-like lines. However,
because they overlap with the absorption lines from the slow
outflow (\S \ref{sec:amd}) they cannot be detected. This in turn
means we underestimate the absorption lines in these troughs.
Kinematically therefore, the narrow emission lines might originate
in the slow outflow. Narrow \x\ emission lines have been
associated with the absorbing outflows in Seyfert galaxies based
on the similar velocities, charge states, and column densities
deduced for the emitting and absorbing plasma
\citep[]{ali02,behar03} . Note that no emission lines from the
fast component of --1900~\kms\ (\S \ref{fast}) are detected. In
fact, strongly blue shifted narrow emission lines are never
detected, neither in the X-ray nor the UV, while slow winds of a
few 100~\kms\ do produce narrow emission lines. Broad (2000~\kms)
emission lines that might be expected if the fast component is
quasi-spherically symmetric are also not observed and in any case
not expected for low charge states, as the fast component is very
highly ionized, as discussed below. All this seems to hint at the
different physical nature (e.g., opening angle and mass) of fast
and slow outflows.

The non-shifted positions ($\Delta v < 70$~\kms) and widths
(\fwhm\ $< 250$ \kms) of the \x\ narrow emission lines are also
consistent with that of the bright, forbidden O$^{+2}$ optical
narrow emission lines at 4959~\AA\ and 5007~\AA, suggesting that
perhaps the \x\ line emitting region is in the optical narrow line
region (NLR). The higher ionization optical (coronal) lines of
ionized Fe appear to be much broader \citep[\fwhm\ $\approx
2000$~\kms,][]{reynolds97}, placing them closer to the nucleus.
However, one has to wonder how robust these widths really are,
given how faint these lines are in \mcg\ \citep[see Fig.~2
in][]{reynolds97}.

\section{RESULTS}
\subsection{Ionic Column Densities}\label{sec:results}

The best-fit ionic column densities are listed in
Table~\ref{table4} and the resulting model is plotted over the
data in Fig.~\ref{figure1}. The Errors for the ionic column
densities were calculated in the same manner as in
\citet{tomer07}. For the most part, the column densities in the
slow component ($-100$~\kms) of the Fe, Si, N, Ne, and Mg ions are
of the order of 10$^{16} - 10^{17}$~cm$^{-2}$, while those of the
more abundant O ions are higher and reach
$\sim$~10$^{18}$~cm$^{-2}$. Comparing our results with those of
\citet{sako03}, we find that Fe L-shell, Si K-shell, Mg K-shell,
Ne K-shell, and N K-shell column densities are more or less
consistent. However, our oxygen ionic column densities are higher
than those of \citet{sako03}. We suspect this may be due to the
better sensitivity to weak absorption lines in the particularly
high signal to noise ratio (S/N) of the present spectrum.
\citet{lee01} obtained still higher O$^{+5}$ and O$^{+6}$ column
densities. This could be a consequence of their need to fit the
17.5~\AA\ drop with an O$^{+6}$ edge. For similar reasons, we
obtain a lower neutral Fe column density than \citet{lee01}. The
O$^{+7}$ column density of \citet{lee01}, on the other hand, is
comparable to the present value.

More significant differences occur for the fast component
($-1900$~\kms). The current Fe K-shell column densities are
slightly higher, but still consistent with those of
\citet{young05}, who used the exact same data set. On the other
hand, \citet{sako03} using the \rgs\ found a fast absorption
component for the Fe L-shell, Ne K-shell, Mg$^{+10}$, Si$^{+12}$,
and O ions, where we find only a slow component. We find the fast
component exclusively in very high ionization species. For O, only
O$^{+7}$ has a fast component. What may have been identified by
\citet{sako03} as high-velocity, low-ionization O, we ascribe to
local ($z = 0$) intervening gas (see \S \ref{local}), whose
apparent velocity in the reference frame of \mcg\ ($z = 0.007749$)
would be --2320~\kms. This velocity is sufficiently close to that
of the fast absorber (--1900~\kms) for the two systems to be
confused by the \rgs, which has lower resolving power than the
presently used \hetgs. The origin and location of the fast,
high-ionization component is further discussed in \S \ref{fast}.

\subsection{$AMD$ For The Slow Component}
\label{sec:amd}

\begin{figure}

\centerline{\includegraphics[width=13cm,angle=0]{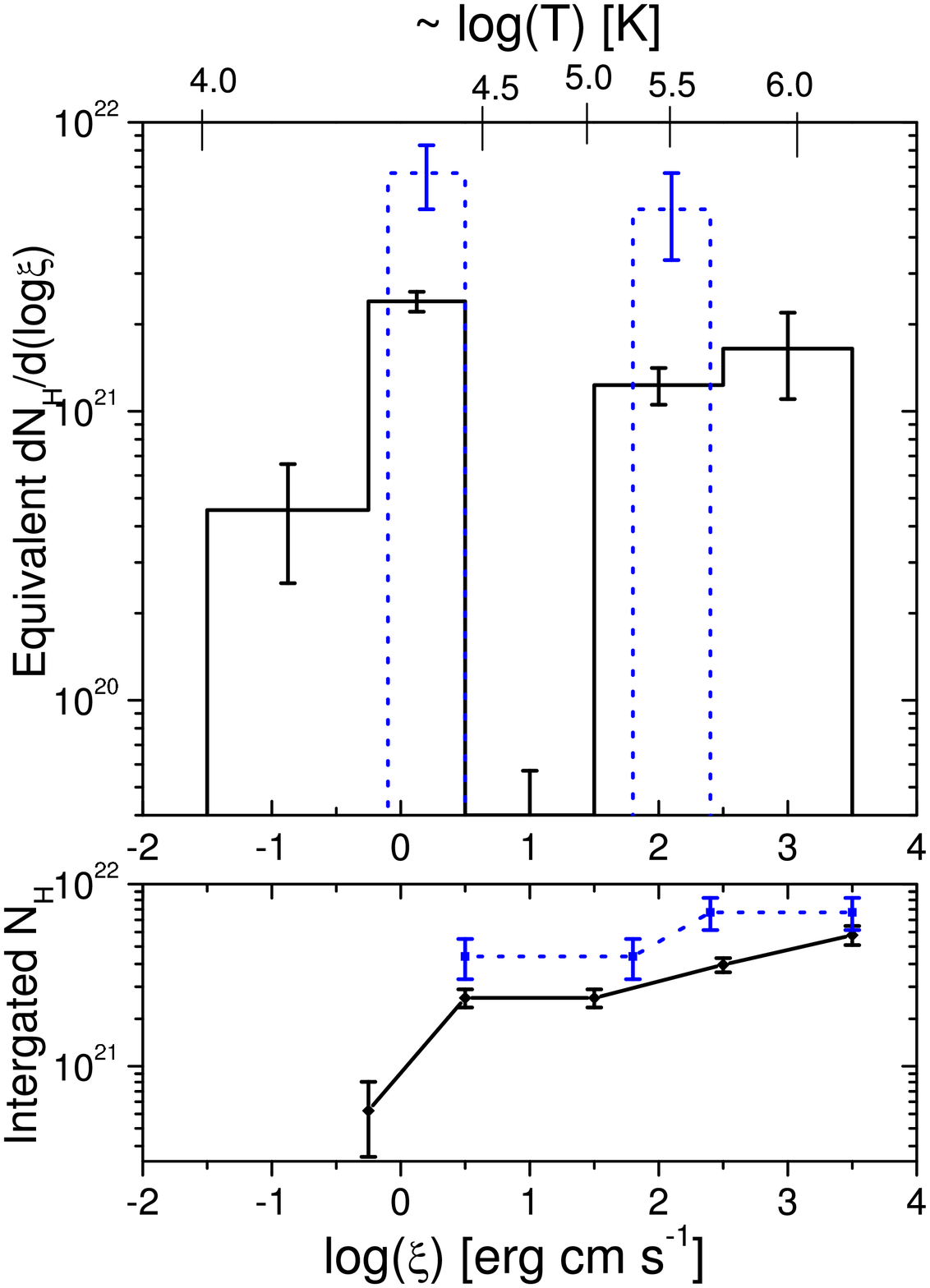}}
\caption{$AMD$ of the slow outflow in \mcg\ obtained exclusively
from Fe absorption and scaled by the solar Fe/H abundance
3.16$\times10^{-5}$ \citep{as09}. The corresponding temperature
scale, obtained from the \xstar\ computation is shown at the top
of the figure. The middle bin value is zero, and only the upper
limit uncertainty is plotted. The cumulative column density up to
$\xi$ is plotted in the lower panel, yielding a total of $N_H~=
(5.3~\pm\ 0.7) \times ~10^{21}$~\cmsq . (Dotted) blue lines
represent the two ionization components of \citet{mckernan07}
broadened by their 3$\sigma$ uncertainty of $\Delta \xi$~= 0.3
\cgs . The cumulative column density of \citet{mckernan07} is
plotted as a blue line in the lower panel, yielding a total of
$N_H~= (7.0~\pm\ 1.4) \times ~10^{21}$~\cmsq. } \label{figure2}
\end{figure}

The best-fit $AMD$ for the slow $-100$~\kms\ absorber in \mcg\ is
presented in Fig.~\ref{figure2} and the integrated column density
is presented in the bottom panel of Fig.~\ref{figure2}. This $AMD$
was obtained using all of the 24 charge states of Fe from neutral
through Fe$^{+23}$. K-shell Fe is not observed for the slow
component and many M-shell ions are only tentatively detected. The
$AMD$ features a statistically significant minimum at $0.5 <
\log\xi < 1.5$ (\cgs), which corresponds to temperatures $4.5 <
\log T <5$ (K). A similar minimum at the same temperatures was
also observed in \iras , \ngc\ \citep{tomer07} and NGC\,7469
\citep{blustin07}. It is mostly a manifestation of the relatively
low ionic column densities observed for the ions
Fe$^{+11}$--~Fe$^{+15}$, as can be seen in Table~\ref{table4}. One
way to explain this gap is that this temperature regime is
thermally unstable \citep{tomer07}. Gas at $4.5 < \log T < 5$ (K)
could be unstable as the cooling function $\Lambda(T)$ generally
decreases with temperature in this regime
\citep[e.g.,][]{krolik81}. Such instabilities could result in a
multi phase (hot and cold) plasma in pressure equilibrium, as
suggested by \citet{krolik81}, and as recently modelled in detail
by \citet{anabella07}. Alternatively, the two distinct ionization
regimes can be ascribed to two geometrically distinct regions
along the line of sight, a high ionization region and a low
ionization region, both which have their own narrow $AMD$
distribution (i.e., well defined $\xi$). However, the fact that
both components appear to have the same outflow and turbulent
velocities leads us to prefer the co-spatial two-phase picture at
the moment. Note that \mcg\ has an even more highly ionized
component, but with a significantly higher outflow velocity of
--1900~\kms . This component probably does come from another
region in the \agn\ and is discussed further in \S \ref{fast}.

The putatively unstable region is also avoided by the
two-component model of \citet{mckernan07}. \citet{sako03} actually
do find ions between $0.5 < \log\xi < 2$ (\cgs). However, their
model does not necessarily preclude a gap within that range. The
model of \citet{lee01} has the low ionization component of the
slow wind $\log\xi$=0.7 \cgs , which on the face of it falls in
the unstable gap. However, that model was calculated with the
\cloudy\ code, while both the present work and \citet{mckernan07}
use \xstar. The unstable region, is known to be very sensitive to
the atomic data and plasma conditions
\citep{hess97,chakravorty09}. The two ionization components of
\citet{mckernan07} are plotted in Fig.~\ref{figure2} over the
presently derived $AMD$. That model of course does not provide a
distribution, but rather two $\xi$ components. Nevertheless, for
the purpose of the plot, we ascribe widths of 0.3 \cgs\ to those
components, which are the 3$\sigma$ quoted errors on the $\log
\xi$ values. It can be seen that the model of \citet{mckernan07},
or any other two component model for that matter, can account for
some of the $AMD$ distribution, but clearly does not realize the
full range of ionization. The {\it total} column densities $N_H$
from both approaches (bottom panel of Fig.~\ref{figure2}) are
formally in agreement, although the two-component model tends to
overestimate the $N_H$ as it needs to produce sufficient ion
abundances far from their maximum-formation temperatures. The
current integrated $AMD$ of the absorber in \mcg\
(Fig.~\ref{figure2}) gives a total column density of $N_H~= (5.3
\pm ~0.7)~\times 10^{21}$~\cmsq , compared with (7.0 $\pm$
1.4)~$\times10^{21}$~\cmsq\ of \citet{mckernan07}.

In order to further compare our results with previous outflow
models for \mcg, we can formally rebin the $AMD$ in
Fig.~\ref{figure2} to two regions, one below ($\log \xi < 0.5$)
and one above ($\log \xi > 1.5$) the thermal instability. The
physical parameters of these two ionization regions are
subsequently compared with all the other works in
Table~\ref{table5}. It can be seen that all of the early works
\citep{otani96, reynolds97, lee01} obtain much too high column
density as they require the model to produce the sharp 17.5~\AA\
turnover with an oxygen absorption edge. None of the models
account for the full range of ionization as the $AMD$ does.

It should be stressed that there are high uncertainties in the
formation temperatures (and $\xi$) of the Fe M-shell ions due to
significant uncertainties in their dielectronic recombination
rates \citep[]{netzer04,badnell06}, as well as the uncertainties
of the EUV and UV ionizing continuum. This could affect the actual
shape of the $AMD$ for $\log \xi < 0.5$~\cgs. The observed minimum
or two-phase structure may consequently change slightly.

\subsection{Fast High-Ionization Component}
\label{fast}

\mcg\ shows two distinct velocity components. The slow one at
$-100$~\kms\ is more prevalent in the spectrum. Only eight ions
are identified for the $-1900$~\kms\ fast component, namely
Fe$^{+23}$, Fe$^{+24}$, Fe$^{+25}$, O$^{+7}$, Mg$^{+11}$,
Si$^{+13}$, S$^{+15}$, and Ar$^{+17}$. See Table \ref{table4} for
their ionic column densities. This fast highly-ionized component
should not be confused with the high-ionization tail of the slow
component. The $AMD$ analysis presented in \S \ref{sec:$AMD$}
refers entirely to the slow component, where both high and low
ionization states are present. An $AMD$ analysis for the fast
component is not possible with only three Fe ions. In fact, it
seems this entire component can be modeled with a single $\xi$
value and total $N_H$. Indeed, we find that log$\xi$ = 3.82~$\pm$
0.03~\cgs\ and $N_H$ = 8.1~$\pm$ 0.7~$\times$10$^{22}$~\cmsq\
yield the measured ionic column densities of Fe$^{+23}$ --
Fe$^{+25}$ to within 10$\%$ and those of other elements to within
60$\%$ for O$^{+7}$, 25$\%$ for Si$^{+13}$, 40$\%$ for S$^{+15}$,
and a factor of 4 for Mg$^{+11}$ and Ar$^{+17}$. We use here a
turbulent velocity of $v_\mathrm{turb}$ = 500~\kms, which is
resolved by \hetgs\ only for $\lambda > 8$~\AA. At shorter
wavelengths, the resolving power of \hetgs\ decreases, as $\Delta
\lambda = 23$~m\AA\ (\fwhm) is fixed. For consistency, we use
500~\kms\ for the entire fast component, which provides a good fit
and reproduces all ionic column densities with a single $\xi$
value. The above quoted errors on $\xi$ and on $N_H$ come from the
distribution of $N_\mathrm{ion}$ values for Fe$^{+23}$ --
Fe$^{+25}$ derived from the single best fit $\xi$. The physical
parameters of this fast component as well as a comparison with
previous works are given in Table~\ref{table6}. Only those works
that could identify and resolve the fast component with gratings
are quoted.

\begin{figure}

\centerline{\includegraphics[width=11cm,angle=90]{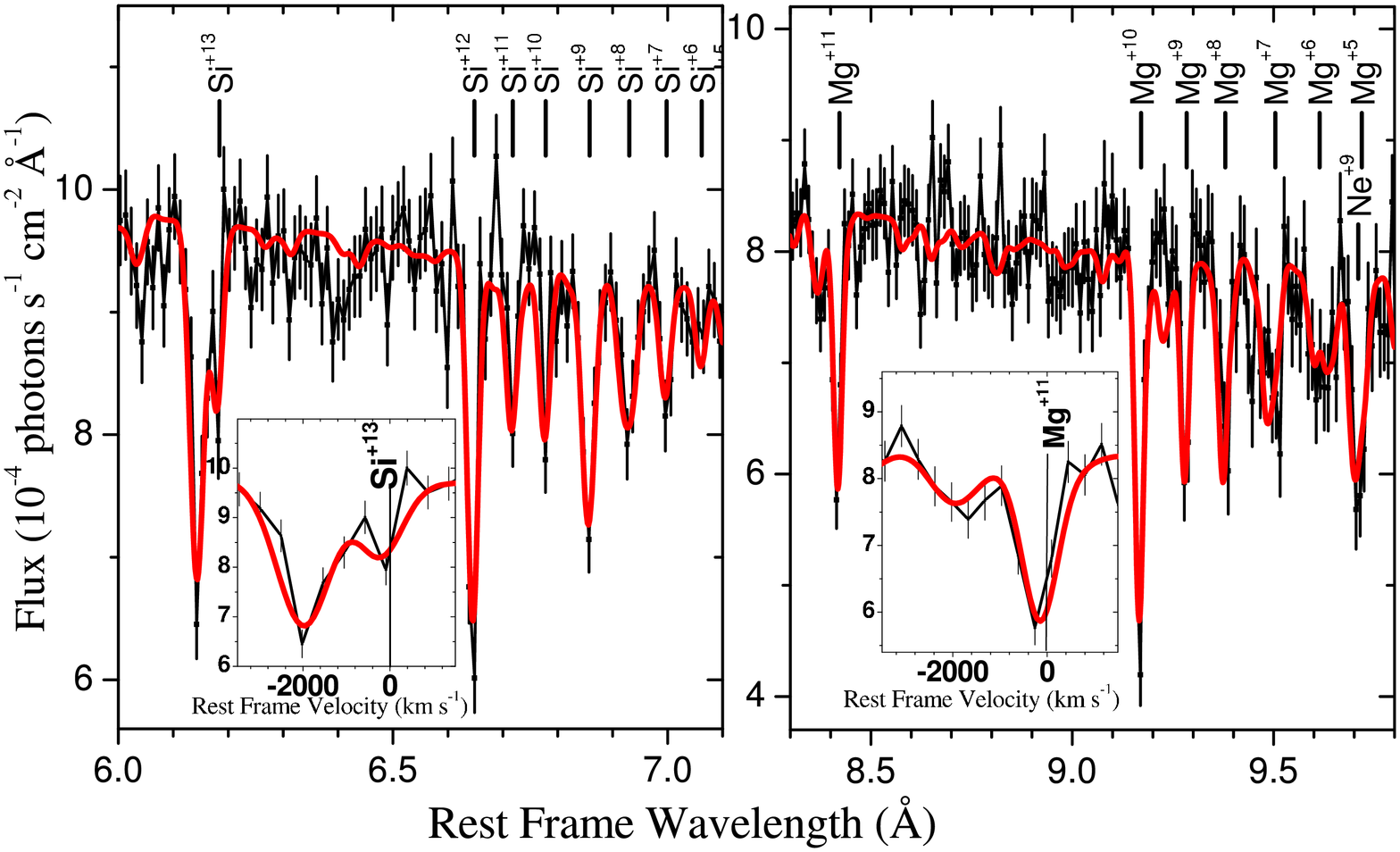}}
\caption{ Extract from the \hetgs\ spectrum of \mcg\ corrected for
cosmological redshift ($z$ = 0.007749). The red line is the
best-fit model including the slow (--100~\kms) and fast
(--1900~\kms) outflow components. ({\it left}) Si$^{+13}$ through
Si$^{+6}$ absorption. Only H-like Si$^{+13}$ shows a fast
component. ({\it right}) Mg$^{+11}$ through Mg$^{+5}$ absorption.
H-like Mg$^{+11}$ shows a fast component, but a much more
prominent slow component. Lower charge states show only the slow
component. The insets show the two H-like lines in velocity space.
The plot demonstrates the low ionization cutoff in the fast
component, which is manifested by only the most ionized species. }
\label{figure3}
\end{figure}

It can be seen in Table~\ref{table6} that all authors more or less
agree on the outflow velocity, although \citet{mckernan07} quote a
somewhat lower value. The turbulent velocity is a less obvious
parameter, but as we argue above it cannot be much lower than what
we use, namely $v_\mathrm{turb}~= 500$~\kms. Not all authors use
such a high turbulent velocity. The column density we obtain is
somewhat higher than in the other works, even though it is still
formally consistent with the results of \citet{young05} and
\citet{mckernan07}. The present value of $\log \xi$~= 3.82 \cgs\
is in good agreement with those of \citet{young05, mckernan07,
miller08}. \citet{sako03} and \citet{chelouche08} claimed to
observe Fe L-shell ions as well as other low ionization species in
the fast component (see Table~\ref{table4}), which resulted in
their lower $\xi$ values, but we conclude that the ionization
parameter of this component needs to be high, and no low
ionization lines exist for it. The exclusively high-ionization
state of the fast component and the contrasting broad ionization
distribution (including very low, see $AMD$ analysis in \S
\ref{sec:amd}) of the slow component are best demonstrated by the
appreciably broad ionization range of inner-shell K$\alpha$
transitions of Si and Mg featured in the compact spectral region
between 6~-- 10~\AA\ \citep{behar02}. This spectral region, which
in \mcg\ comprises the absorption lines of Si$^{+5}$ through
Si$^{+13}$ and of Mg$^{+5}$ through Mg$^{+11}$ is depicted in
Fig.~\ref{figure3}. The lines in this limited waveband cover the
significant ionization range of roughly $-1 < \log \xi < 2.5$
(\cgs). First, it can be seen that both Si$^{+13}$ and Mg$^{+11}$
H-like ions have absorption lines from both the slow and fast
components as manifested by their double troughs. However, while
the fast component is much more prominent for Si$^{+13}$, the
opposite is true for the less ionized Mg$^{+11}$. This is a sign
that the fast component is weaker in the less ionized species.
Second, all of the lower charge states, namely He-like and into
the L-shell, of both elements have only a slow component
absorption line, as can be seen by the proximity of the troughs to
the rest frame wavelengths of their respective transitions labeled
in Fig.~\ref{figure3}.

The outflow velocity of the fast component ($-1900$~\kms) is
close, but significantly different (and clearly resolved by
\hetgs) from the cosmological recession of \mcg\ of $-2320$~\kms,
possibly confusing the fast component with local ionized ISM
absorption. This point was already discussed by \citet{young05},
who showed that not only are the velocities slightly different,
but also the high column density measured in the fast component
would require an ISM absorber three orders of magnitude larger
than the size of our Galaxy. Furthermore, the high ionization of
the fast component up to, e.g., Fe$^{+25}$ and Ar$^{+17}$, is much
higher than typically found in intergalactic absorbers. We
conclude that the fast component is most likely intrinsic to \mcg.
The possible confusion of low-ionization local oxygen lines with
the fast component is further discussed and clarified in \S
\ref{local}.

The well defined ionization parameter found for the fast outflow
suggests it may be described as a uniform, spherical, thin shell.
Its high column density and ionization suggest, in turn, it could
lie rather close to the central \agn\ source. The width of the
shell can be denoted by  $\Delta r = N_H / n_H$, where $n_H$ is
the hydrogen number density. Using the definition of $\xi$, $n_H=L
/ ( \xi r^2)$, and one can write $\Delta r = N_H \xi r^2 / L$.
Requiring now that $\Delta r < r$ leads to an upper limit on the
distance from the center of $r <
10^{16}~L_{43}N_{H23}^{-1}\xi_4^{-1}$~cm, or $r <$~0.0093~pc,
which is about 11 light days. Above, $L_{43}$ is the 1~--
1000~Rydberg luminosity in units of 10$^{43}$~erg~s$^{-1}$,
$\xi_4$ is the ionization parameter in units of 10$^4$ \cgs , and
$N_{H23}$ is the hydrogen column density in units of 10$^{23}$
cm$^{-2}$. The continuum we use in this work yields $L_{43} =
1.5$. Note that the above distance estimate is somewhat affected
by the line velocity broadening used in the model. With a
turbulent velocity lower than 500~\kms, the derived column density
would be slightly higher, and hence the estimated distance would
be slightly lower. However, as argued above, the turbulent
velocity is probably not much lower than 500~\kms. The present
distance of 11 light days for the fast component is roughly the
same as the estimate of \citet[][referred to there as zone
3]{miller08}. For comparison, this is a few times the broad line
region (BLR) distance of \mcg. The BLR distance $R_\mathrm{BLR}$
can be estimated from its correlation \citep[][Fig.~5]{bentz08}
with the 5100~\AA\ luminosity of $\lambda L_\lambda~= 1.4 \times
10^{42}$ erg\,s$^{-1}$ \citep[taken from][e.g.,Fig.~2]{reynolds97}
to be approximately 4 $\pm$ 2 light days. The estimated black hole
mass from the H$\beta$ \fwhm\ of 2400~\kms\ \citep{reynolds97} and
eq. (5) in \citet{kaspi00} is 3$\times 10^6$ M$_{\odot}$ with the
standard factor of 2--3 uncertainty for such estimates. The
Keplerian velocity at the absorber distance of 10 light days,
therefore, is roughly 1200 \kms, which is slightly higher, but
still consistent within the errors with the observed line
broadening of $v_\mathrm{turb} = 500$~\kms\ (\fwhm = 830~\kms).

The mass outflow rate $\dot{M}$ in the fast component can also be
estimated within this thin-shell constant-density approximation.
The product of the particle flux in the wind $n_Hv_\mathrm{out}$,
the cross section of the absorbing shell $\Omega r^2$ assuming
conical geometry, and the average particle mass $\mu m_H$ yields

\begin{equation}
\dot{M}  = n_H v_\mathrm{out} \Omega r^2 \mu m_H =
 \frac{L v_\mathrm{out} \Omega \mu m_H}{\xi}
\label{mdot}
\end{equation}

\noindent where $v_\mathrm{out}$ is the outflow velocity
(--1900~\kms\ here) and $\Omega$ is the unknown opening solid
angle of a presumably conical flow. For the right hand side of
Eq.~(\ref{mdot}), we used as before the expression for the
ionization parameter $\xi = L / (n_Hr^2)$. Plugging in typical
values yields

\begin{equation}
\dot{M} \simeq 0.043 \frac{\Omega_{0.5}L_{43}v_{2000}}{\xi_4}~
M_\bigodot \mathrm{yr}^{-1} \label{mdot2}
\end{equation}

\noindent where $v_{2000}$ is the outflow velocity in units of
2000 \kms, $\Omega_{0.5}$ is the opening solid angle as a fraction
0.5 of $4\pi $~str, and $\mu = 1.3$ is assumed. This mass loss
rate needs to be further suppressed in \agn\ feedback estimates,
if the duty cycle of the flow over time (or effectively the radial
volume filling factor) is substantially less than unity. The mass
loss rate of eq.~(\ref{mdot2}) implies kinetic power of
$\dot{M}v^2/2 = 1.06\times10^{41}$~erg\,s$^{-1}$, which is two
orders of magnitude less than $L_\mathrm{bol} = 8~\times
10^{43}$~erg\,s$^{-1}$ \citep{reynolds97}. The outflow is more
substantial in terms of momentum, as $(\dot{M}v) /
(L_\mathrm{bol}/c) \approx 0.4$. If the outflow is driven solely
by radiation pressure then 80$\%$ of the radiation should be
absorbed by the gas (a factor of 2 comes from the outflow opening
angle compared with the bolometric 4$\pi$ coverage). We find that
the absorbed \x\ flux in our model is $\sim$26$\%$. This value
seems much lower than the 80$\%$ needed. However, the main
radiation driven mechanism occurs in the UV band on which we do
not have reliable data. Therefore, we cannot reach a clear
conclusion, even though effective optical depth of 0.8 seems too
high, at least in the \x\ band. The considerably less ionized
state of the slow wind suggests it may carry much more mass than
the fast component (eq.~\ref{mdot2}). However, the broad $AMD$ of
the slow component necessarily implies different physical
conditions are present and a more complicated geometry than the
simplified uniform, thin-shell picture assumed here for the fast
component.

\subsection{A Third, Local Absorption Component}
\label{local}

In this section, we focus on the long wavelength region of the
spectrum, which is filled predominantly with oxygen lines from
highly ionized species down to neutral. Some of these lines have
been ascribed, we believe erroneously, in previous works
\citep{sako03, chelouche08} to the fast outflow. As demonstrated
in \S \ref{fast}, the absorption signature of the fast outflow at
--1900~\kms\ actually diminishes rapidly with decreasing $\xi$. In
fact, only the eight most highly ionized species show absorption
at --1900~\kms.

\begin{figure}

\centerline{\includegraphics[width=13cm,angle=90]{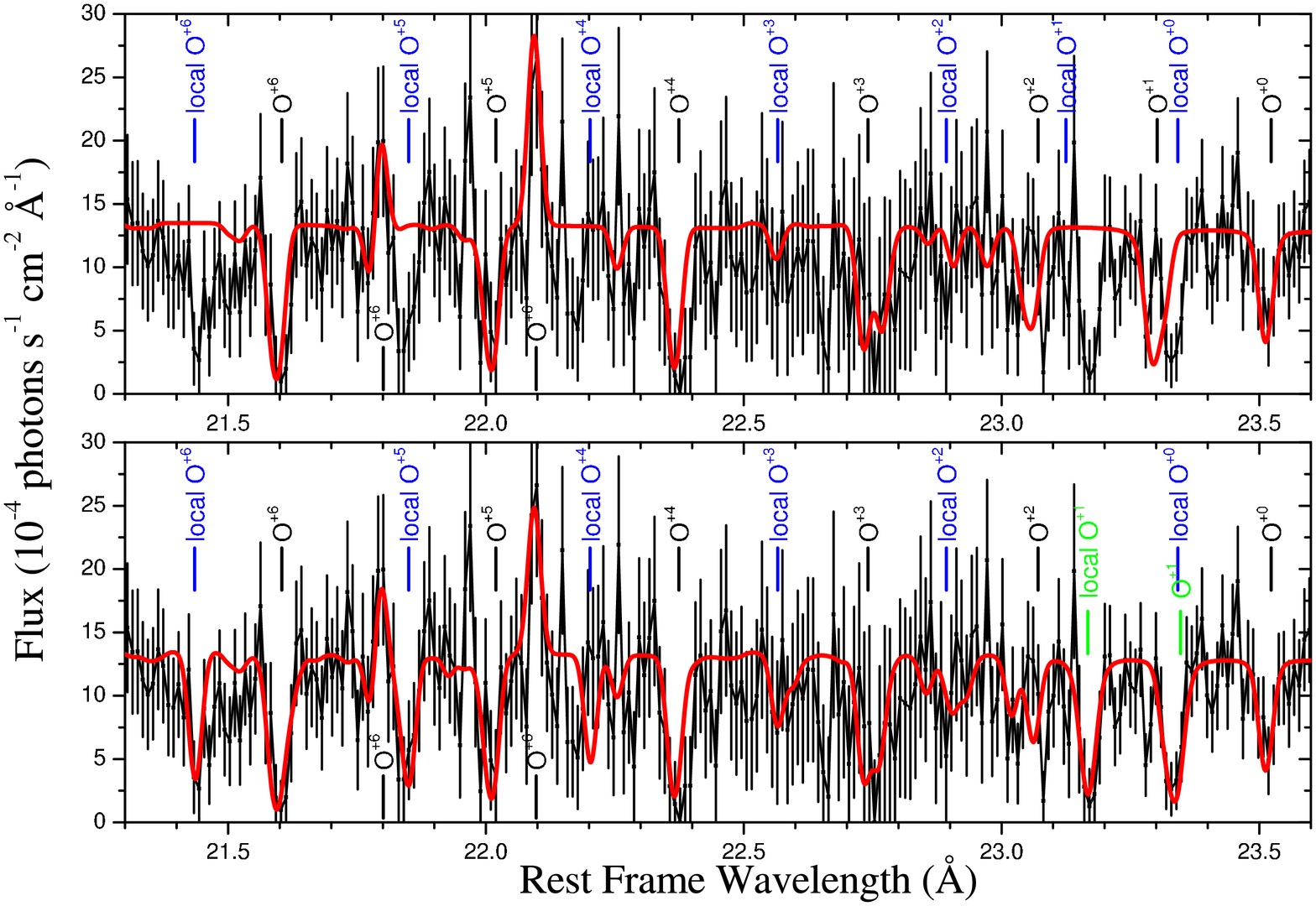}}
\caption{ Oxygen K$\alpha$ line region of the \hetgs\ spectrum of
\mcg\ corrected for cosmological redshift ($z = 0.007749$). Red
curve indicates model. Most important lines are labeled at their
rest frame. Labels "local" refer to positions expected for the $z
= 0$ absorber. Green labels refer to adjusted wavelengths (\S
\ref{corrections}). {\it (top)} Model includes only \agn\ (mostly
slow --100~\kms) outflow. {\it (low)} Improved model includes also
local absorption.} \label{figure4}
\end{figure}

The relevant part of the spectrum that is crowded with lines from
neutral oxygen to O$^{+6}$ is plotted in Fig.~\ref{figure4}.
Indeed, the numerous lines as well as the uncertainty associated
with some of their wavelengths make the analysis of this spectral
region particularly challenging. The top panel of
Fig.~\ref{figure4} shows the model with absorption lines by the
slow outflow (\S \ref{sec:amd}). Due to the low $-100$~\kms\
blueshift, these absorption lines lie just short of their
rest-frame labels in the figure. Systematically blueshifted from
these positions, additional absorption lines, not accounted for by
the model in the upper panel, can be identified. These lines are
shifted by $-2300 \pm 150$~\kms. Above 20~\AA, this shift is
readily discriminated by \hetgs\ from the $-1900~\pm 150$~\kms\
velocity of the fast highly-ionized component. Furthermore, the
oxygen lines in Fig.~\ref{figure4} are narrower than the
relatively broad ($v_\mathrm{turb}$~= 500~\kms) lines of the fast
wind, which further precludes the oxygen lines from pertaining to
the fast wind. On the other hand, a shift of $-2300$~\kms\ is
exactly the cosmological redshift $z = 0.007749$ of \mcg. Note
that the spectrum in Fig.~\ref{figure4} has been de-redshifted to
the \agn\ rest frame (as have been all other spectra plotted in
this paper). Hence, lines that appear blueshifted by $-2300$~\kms\
are actually at rest in the local frame of reference. We therefore
interpret these lines as arising locally from absorption by
ionized ISM in our galaxy or in the Local Group at $z = 0$. The
widths of the local oxygen absorption lines are not resolved here,
but we use $v_\mathrm{turb}$~= 100~\kms\ to model the troughs and
to obtain ionic column densities. This width is consistent with
local, ionized ISM \uv\ absorption lines at high resolution
\citep{shai04}.

The lower panel in Fig.~\ref{figure4} shows the local $z = 0$
absorption component added to the model and it can be seen to
provide a much improved fit and to account for most of the
absorption that was missing in the model in the upper panel. We
note that a few wavelengths had to be slightly corrected from
their calculated value in order to fit the data. Wavelength
adjustments facilitated by the \hetgs\ spectrum are further
discussed in \S \ref{corrections}. An exception to the good fit is
the O$^{+4}$ feature predicted at $\sim ~22.17$~\AA. It appears to
be present in the data, but shifted by $\sim ~0.2$~\AA, or
--300~\kms\ by comparison to the model. We checked whether this
line could be due, alternatively, to absorption from excited,
meta-stable levels of O$^{+4}$ \citep{kaastra04}. However, the
strongest such lines are expected at 22.488 \AA\ and at 22.453
\AA. Although there is some flux deficit in the spectrum around
the latter position, it looks more like noise, and we find nothing
that appears broad enough to be a conclusive absorption line (see
Fig.~\ref{figure4}). We conclude that the feature at 22.17~\AA\
cannot be explained by absorption from meta-stable levels.

\begin{figure}

\centerline{\includegraphics[width=13cm,angle=90]{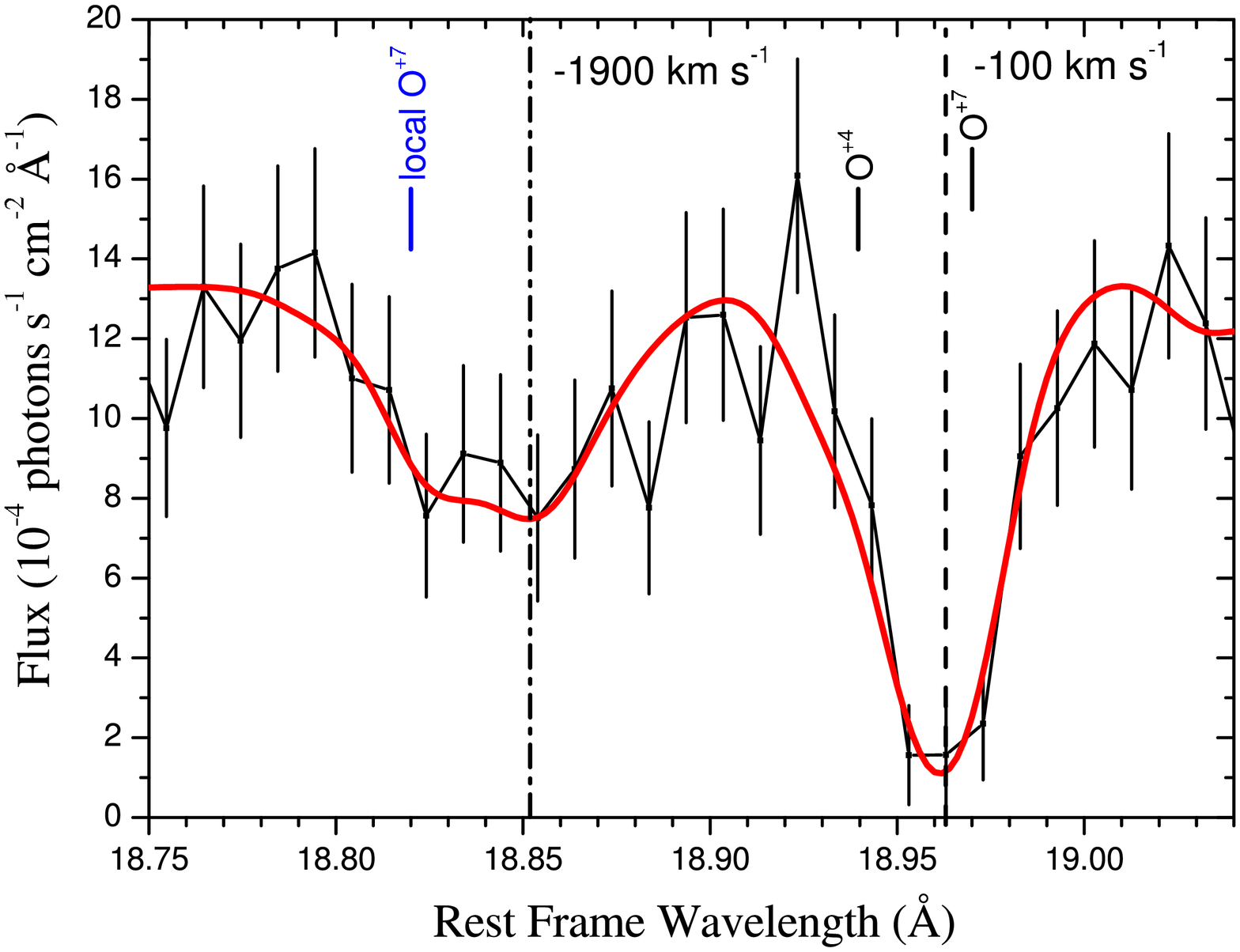}}
\caption{\hetgs\ spectrum of \mcg\ corrected for cosmological
redshift ($z = 0.007749$) around the prominent O$^{+7}$ Ly$\alpha$
line. The red line is the (globally) best-fit model. Absorption by
both slow (--100~\kms) and fast (--1900~\kms) components are
clearly discerned (unlike lower oxygen charge states in
Fig.~\ref{figure4} that have only a slow component). The fast
component is also broader ($v_\mathrm{turb}~= 500$~\kms) than the
unresolved slow one ($v_\mathrm{turb}~= 100$~\kms\ further
broadened by the instrument). The marginally detected
local-component line is also indicated.}
\label{figure5}
\end{figure}

The local column densities obtained from fitting the oxygen lines
(Fig.~\ref{figure4}) are listed in Table~\ref{table7}. The columns
in the (slow) outflow component are also listed for comparison. It
can be seen that neutral O, O$^{+1}$, O$^{+5}$, and O$^{+6}$ are
unambiguously detected. Local absorption by O$^{+2}$ -- O$^{+4}$
and by O$^{+7}$ are less significant. The data around the leading
K$\alpha$ lines of O$^{+2}$ and O$^{+3}$ are noisy
(Fig.~\ref{figure4}). O$^{+4}$ appears to have a considerable
absorption trough, but it does not exactly agree in position with
the rest-frame wavelength. Local absorption in O$^{+7}$
Ly\,$\alpha$ is blended with absorption in the fast component of
that line. Indeed, we can only put an upper limit to the local
component in O$^{+7}$ of 10$^{16}$~\cmsq. In fact, the Ly$\alpha$
line of O$^{+7}$ has a different absorption profile than any other
oxygen line from lower charge states, and is shown separately in
Fig.~\ref{figure5}. It can be clearly seen to have a broad and
fast --1900~\kms\ component that is distinct from the local
component. This result fits well with our assessment of the fast
outflow being exclusively comprised of high ionization species,
while the slow outflow and the local absorber are significantly
less ionized. {\it IUE} spectra of \mcg\ show no flux below
3000$\AA$ which prohibits detection of oxygen lines in the UV.

Applying an ionization correction for the fractional abundances of
$f_q \approx 0.5$ to the highest column-density ion O$^{+1}$ for
which $N_{ion} = 10^{17}$~\cmsq , and then dividing by the solar
O/H abundance (4.6$\times$10$^{-4}$) yields a rough estimate to
the equivalent hydrogen column density of $\sim
4\times10^{20}$~\cmsq, which is the same as the neutral column
towards \mcg, and for which our measurements now reveal an ionized
phase. We can conclude there is roughly the same amount of ionized
oxygen as neutral oxygen in the direction of \mcg.
Table~\ref{table7} shows that while the slow component has very
high column densities in the high charge states O$^{+5}$ --
O$^{+7}$, the column distribution of the local absorber is rather
flat. The high intrinsic outflow columns are manifested in the
prominent absorption from these ions, even in lines from
relatively weak high-order transitions (Fig.~\ref{figure6}).
Conversely, high-order lines from the local component are
inconspicuous and hardly detected. Spectral features due to dust
grains with oxide composites may also be present in the data, but
only to the extent that can explain the residuals to the atomic
model plotted in Fig.~\ref{figure4}.

\section{Oxygen Wavelength Adjustments}
\label{corrections}

Wavelengths of inner-shell absorption lines are obtained primarily
from atomic computations and are difficult to benchmark in the
laboratory. Two notable laboratory measurements relevant to
inner-shell oxygen ions, albeit in emission, were published by
\citet{schmidt04} and \citet{gu05}. These measurements provide
useful wavelengths of blends, even though some individual
absorption lines remain hard to discern. The present \hetgs\
spectrum of \mcg\ is of sufficiently high spectral resolution and
S/N to be directly confronted with the computed wavelengths.

\begin{figure}

\centerline{\includegraphics[width=13cm,angle=90]{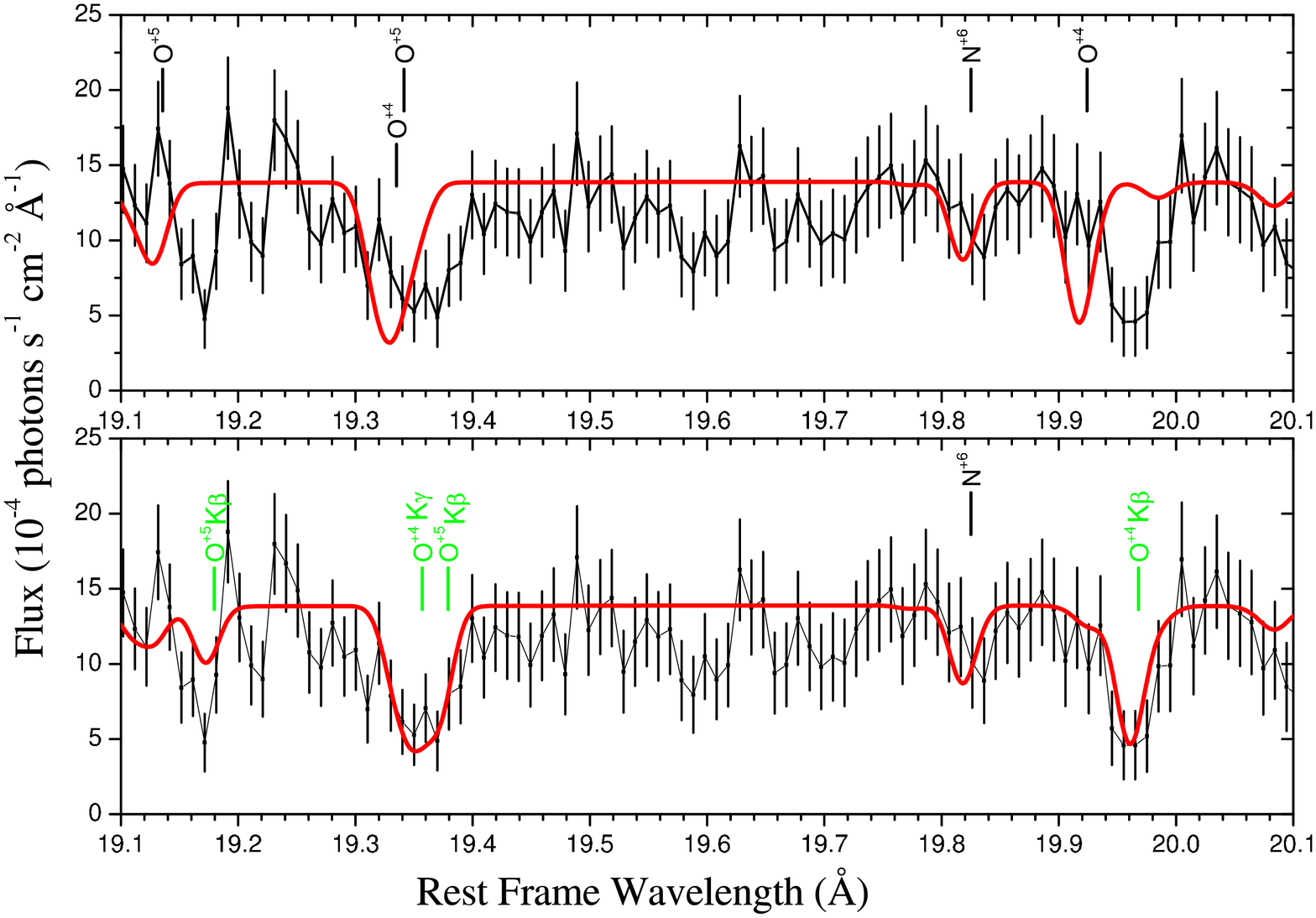}}
\caption{ Extract from \hetgs\ spectrum of \mcg\ corrected for
cosmological redshift ($z$ = 0.007749) demonstrating the ability
of \hetgs\ to accurately calibrate rest-frame wavelengths. The red
lines represent the model with the --100~\kms\ blueshift of the
slow wind. ({\it top}) Model using \hullac\ computed wavelengths.
({\it bottom}) Improved model using re-calibrated wavelengths.}
\label{figure6}
\end{figure}

The published wavelengths for the leading K$\alpha$ absorption
lines of O$^{+4}$ and O$^{+5}$ from laboratory measurements are
22.374~\AA\ and 22.019~\AA, respectively. Indeed, these
wavelengths give a very good fit to the data with the outflow
velocity of --100~\kms, which is prevalent throughout the slow
outflow component (see Fig.~\ref{figure4}). On the other hand, the
computed positions of the corresponding higher order lines
(K$\beta$, K$\gamma$, etc.) show slight discrepancies when
compared with the observed absorption lines, as demonstrated in
the upper panel of Fig.~\ref{figure6}. This inconsistency can be
remedied by adjusting the computed yet uncertain rest-frame
wavelengths of these lines to match the observed absorption lines.
The good agreement found for the leading K$\alpha$ lines makes the
high-order wavelength adjustments independent of any kinematic
uncertainty, since all lines of a given ion must be Doppler
shifted by the exact same velocity. We invoke corrections of up to
45~m\AA\ to the computed wavelengths, which is comparable in
magnitude to the maximal discrepancies found between \hullac\ and
measured wavelengths of the leading K$\alpha$ lines in laboratory
measurements \citep{schmidt04, gu05}. The complete list of
adjusted wavelengths is presented in Table~\ref{table8}. The lower
panel in Fig.~\ref{figure6} shows the best-fit model following the
wavelength adjustments, which is clearly favored by the data. Note
that the absorption line strengths (i.e., equivalent widths) have
not changed much between the two panels of Fig.~\ref{figure6}, as
the ionic column densities are essentially anchored by the leading
K$\alpha$ transitions (shown in Fig.~\ref{figure4}).

Based on the observed spectrum around 23~\AA\
(Fig.~\ref{figure4}), we also inspect the leading K$\alpha$ lines
and blends of O$^{+1}$~-- O$^{+3}$. The O$^{+1}$ lines are better
constrained by the deep trough of the local component, while the
slow outflow component of this ion blends with local absorption by
neutral O (see Fig.~\ref{figure4}). The three strongest lines of
O$^{+1}$ are unresolved in the spectrum and require a uniform
shift of +45~\AA, which nicely produces the observed O$^{+1}$
absorption trough. The significant improvement of the model
following the wavelength adjustments is demonstrated in the bottom
panel of Fig.~\ref{figure4}. The O$^{+2}$ and O$^{+3}$ lines, both
in the outflow and locally are rather weak in the spectrum.
Consequently, although the computed wavelengths are likely
somewhat inaccurate, reliable adjustments for these lines are
unwarranted by the data. We set therefore the strongest O$^{+2}$
line at 23.071~\AA, which is the position of an (emission) line
blend identified by \citet{gu05}, and uniformly shift weaker
\hullac\ lines by the same amount (Table~\ref{table8}). For
O$^{+3}$, we set the leading blend to 22.741~\AA\ \citep[again, an
emission blend in][]{gu05} and shift the other lines by the same
amount with respect to their \hullac\ positions. The summary of
all adjusted wavelengths is given in Table~\ref{table8}.

The above improved wavelengths obtained from the \hetgs\ spectrum
of \mcg\ can be used for better line identification in other
astrophysical absorption spectra. However, given the S/N and the
uncertain conditions at the source, the new wavelengths should be
trusted to no more than $\pm 10$~m\AA. This calibration of
wavelengths with \hetgs\ can be viewed as a form of laboratory
astrophysics from space. However, by no means can it replace
actual absorption measurements carried out under controlled
conditions in the laboratory.

\section{CONCLUSIONS}
\label{sec:concl}

We have analyzed the kinematic and thermal structure of the
ionized outflow in \mcg. We find three distinct absorption
systems, two of which are intrinsic to the \agn. The slow
component is outflowing at --100~\kms\ and spans a considerable
range of ionization from neutral Fe to Fe$^{+23}$ ($-1.5 < \log
\xi< 3.5$~\cgs). A second, fast outflow component at --1900~\kms\
is very highly ionized ($\log \xi = 3.8$~\cgs). Finally, a third
component of local absorption at $z = 0$ is detected for the first
time in \mcg\ by its oxygen absorption.

Using our $AMD$ reconstruction method for the slow component, we
measured the distribution of column density as a function of
$\xi$. We find a double-peaked distribution with a significant
minimum at $0.5 < \log \xi < 1.5$ (\cgs) , which corresponds to
temperatures of $4.5 < \log T < 5$ (K). This minimum was observed
in several other \agn\ outflows and it can be ascribed to thermal
instability that appear to exist ubiquitously in photo-ionized
Seyfert winds. The fast outflow with its narrow ionization
distribution can be described as a thin shell and estimated to be
approximately ten light days away from the central ionizing
source. The local absorption system we believe could arise from
either the ionized Galactic ISM, or from the Local Group. Finally,
we use the \hetgs\ spectrum to slightly improve on the computed
wavelengths of the most important inner-shell O lines.

\begin{acknowledgements}
We thank Shai Kaspi and Doron Chelouche for useful discussions. TH
and NA acknowledge support from \chandra\ grant AR-7 8011A. EB
acknowledges funding from NASA grant 08-ADP08-0076.
\end{acknowledgements}

\clearpage

\clearpage

\clearpage

\clearpage

\clearpage

\clearpage

\clearpage

\begin{deluxetable}{lccccc}
\tablecolumns{4} \tablewidth{0pt} \tablecaption{Historical
observations of the ionized absorber in \mcg\ \label{table1}}
\tablehead{
   \colhead{Observatory } &
   \colhead{Year } &
   \colhead{Duration } &
   \colhead{Flux level at 1 keV } &
   \colhead{References \tablenotemark{a}} \\
   \colhead{}&
   \colhead{} &
   \colhead{(ks)} &
   \colhead{(10$^{-3}$ ph cm$^{-2}$ s$^{-1}$ keV$^{-1}$ ) } &
   \colhead{}}

\startdata
\rosat & 1992  & 8.5 & 6 &1   \\
\asca & 1994  & 150  & 9 & 1, 2       \\
\chandra &2000 & 120 & 10 & 4, 7   \\
\xmm  & 2000 & 120 & 6 & 3, 5, 8   \\
\xmm & 2001 & 320 & 9 &  6, 8 \\
\chandra  &2004 & 520 & 10 & 8, 9, 10  \\
\suzaku &2006 & 250 & 8  & 8

\enddata

\footnotesize
\tablenotetext{a}{ 1. \citet{reynolds97}, 2. \citet{otani96}, 3. \citet{brand01}, 4. \citet{lee01}, 5. \citet{sako03}, 6. \citet{turner04}, 7. \citet{mckernan07}, 8. \citet{miller08}, 9. \citet{chelouche08}, 10. present work.}

\end{deluxetable}

\begin{deluxetable}{lcccccc}

\tablecolumns{6} \tablewidth{0pt} \tablecaption{\chandra\
observations of \mcg\ used in this work \label{table2}}
\tablehead{

   \colhead{Obs. ID} &
   \colhead{Start} &
   \colhead{Detector} &
   \colhead{Gratings} &
   \colhead{Exposure} &
   \colhead{Counts in HEG} &
   \colhead{Counts in MEG } \\

   \colhead{ }&
   \colhead{Date} &
   \colhead{} &
   \colhead{} &
   \colhead{(s) } &
   \colhead{orders $\pm$ 1} &
   \colhead{orders $\pm$ 1 }
}

\startdata
4760 & 2004 May 19 & ACIS-S  & HETG &169590 & 50630 & 107040  \\
4761 & 2004 May 21& ACIS-S & HETG &156230 & 51279 & 109923\\
4759 &  2004 May 24& ACIS-S  & HETG & 158535 & 50152 &108233 \\
4762 & 2004 May 27& ACIS-S & HETG &37549 &13637  &28289  \\
\enddata

\end{deluxetable}

\begin{deluxetable}{lccc}
 \tablecolumns{3} \tablewidth{0pt}
 \tablecaption{Narrow Emission
Lines }

\tablehead{

   \colhead{Line} &
   \colhead{$\lambda _\mathrm{Rest}$} &
   \colhead{$\lambda _\mathrm{Observed}$ \tablenotemark{a} }&
   \colhead{Flux} \\
   \colhead{ }&
   \colhead{(\AA)} &
   \colhead{(\AA)}  &
   \colhead{(10$^{-5}$ photons s$^{-1}$ cm$^{-2}$) }
}

\startdata

Fe$^{+0}$ -- Fe$^{+9}$ K$\alpha$   & 1.94 &  \multicolumn{1}{c}{\multirow{2}{*}{1.940~$\pm$ 0.006~\tablenotemark{b}}} &  \multicolumn{1}{c}{\multirow{2}{*}{ 1.0~$\pm$ 0.2}}  \\
Fe$^{+10}$ -- Fe$^{+16}$ K$\alpha$   & 1.93 -- 1.94~\tablenotemark{d}   &   &  \\
\multicolumn{1}{l}{\multirow{2}{*}{Fe$^{+17}$ -- Fe$^{+23}$ K$\alpha$}}& \multicolumn{1}{l}{\multirow{2}{*}{1.86 -- 1.90~\tablenotemark{e}}} &  1.905~$\pm$ 0.006~\tablenotemark{b} &  0.5~$\pm$ 0.1  \\
                                   &                                &  1.877~$\pm$ 0.006~\tablenotemark{b} &  0.4~$\pm$ 0.1  \\
Ne$^{+8}$ forbidden &  13.698 & 13.710~$\pm$ 0.012~\tablenotemark{c} & 0.5 $\pm$ 0.1   \\
O$^{+6}$ intercombination&  21.801  & 21.794~$\pm$ 0.009~\tablenotemark{c}& 3.5 $\pm$ 0.6\\
O$^{+6}$ forbidden& 22.097  & 22.093~$\pm$ 0.009~\tablenotemark{c}
& 6~$\pm$ 1 \label{table3}
\enddata

 \footnotesize
\tablenotetext{a}{ in the \agn\ rest frame.}
 \tablenotetext{b}{ FWHM = 15 m\AA.}
 \tablenotetext{c}{ FWHM = 235 \kms .}
 \tablenotetext{d}{ \citet{decaux95}.}
\tablenotetext{e}{ \citet{decaux97}.}

\end{deluxetable}

\begin{deluxetable}{lccccccc}
\tabletypesize{\footnotesize} \tablecolumns{6} \tablewidth{0pt}
\tablecaption{Current best-fit
 column densities for ions detected in the 2004 \hetgs\ spectrum
 of \mcg\ compared with the 2000 \rgs\ spectrum \citep{sako03}.
 \label{table4}} \tablehead{
   \colhead{Ion} &
   \colhead{\hetgs\  } &

   \colhead{\rgs\ } &
   \colhead{Ion} &
   \colhead{\hetgs\   } &

   \colhead{\rgs\  } \\

   \colhead{} &
   \colhead{$N_{\mathrm ion}$} &

   \colhead{$N_{\mathrm ion}$} &
   \colhead{} &
   \colhead{$N_{\mathrm ion}$} &

   \colhead{$N_{\mathrm ion}$} \\

   \colhead{} &
   \colhead{(10$^{16}$~cm$^{-2}$)} &

   \colhead{(10$^{16}$~cm$^{-2}$)} &
   \colhead{} &
   \colhead{(10$^{16}$~cm$^{-2}$)} &

   \colhead{(10$^{16}$~cm$^{-2}$)}
}
  \startdata
&    {Slow~~Fast} &   {Slow~~Fast} & &  {Slow~~Fast} & {Slow~~Fast}  \\
\cline{2-3} \cline{5-6} N$^{+6}$    &
$10_{-3}^{+8}$~~~~\nodata     &   $3.7$~~~~$1.3$                                      & Si$^{+11}$  &   $2.0_{-1.5}^{+0.2}$~~~~\nodata &   \nodata                                    \\
\cline{1-3}
O$^{+0}$    &  $10_{-5}^{+5}$~~~~\nodata &\nodata                                           & Si$^{+12}$  &   $12.0_{-3.5}^{+1.5}$~~~~\nodata  &   $\leq12.6$~\tablenotemark{a}~~~~~~$3.3$   \\
O$^{+1}$    &  $8_{-1}^{+16}$~~~~\nodata &\nodata                                           &             Si$^{+13}$& $3.0_{-1.0}^{+0.5}$~~~~$8_{-1}^{+0.5}$  &$\leq10.7$~\tablenotemark{a}~~~$\leq34.7$~\tablenotemark{a}   \\
\cline{4-6}
O$^{+2}$    &  $1.0_{-0.5}^{+2.0}$~~~~\nodata &\nodata                                           &   S$^{+15}$            &  \nodata~~~~$6_{-1.1}^{+5.4}$  & \nodata   \\
\cline{4-6}
O$^{+3}$    &  $6_{-0.6}^{+12}$~~~~\nodata    &   \nodata                                   &   Ar$^{+17}$            &  \nodata~~~~$8_{-2.4}^{+3.2}$  & \nodata\\
\cline{4-6}
O$^{+4}$    &$12_{-5}^{+3}$~~~~\nodata &   $8.7$~~$\leq1.6$~\tablenotemark{a}         &   Fe$^{+0}~\tablenotemark{e}$   &  $5_{-0.6}^{+1}$~~~~\nodata &   \nodata\\
O$^{+5}~\tablenotemark{b}$    &   $12_{-6}^{+1.5}$~~~~\nodata  &  $3.6$~~~~$1.0$             & Fe$^{+1}$   &   $4_{-0.8}^{+1}$~~~~\nodata  &   \nodata \\
O$^{+6}~\tablenotemark{c}$    &   $60_{-6}^{+7.5}$~~~~\nodata &  $22$~~~~$1.4$               &  Fe$^{+2}$   &   $0.5_{-0.5}^{+0.3}$~~~~\nodata  &   \nodata \\
O$^{+7}~\tablenotemark{d}$    &   $120_{-24}^{+12}$~~~~$2_{-0.3}^{+1.6}$   &   $19$~~~~$5.6$          &  Fe$^{+3}$   &   $0.5_{-0.5}^{+0.5}$~~~~\nodata  &   \nodata\\
\cline{1-3}
Ne$^{+3}$   &   $2_{-0.8}^{+5}$~~~~\nodata &   \nodata                                      & Fe$^{+4}$   &   $0.6_{-0.6}^{+0.2}$~~~~\nodata  &   \nodata  \\
Ne$^{+4}$   &   $4_{-0.6}^{+3.4}$~~~~\nodata &   \nodata                                      &  Fe$^{+5}$   &  $0.7_{-0.7}^{+0.2}$~~~~\nodata   &   \nodata \\
Ne$^{+5}$   &   $4_{-0.6}^{+6}$~~~~\nodata &   \nodata                                     &  Fe$^{+6}$   &   $0.7_{-0.7}^{+0.2}$~~~~\nodata  &    \nodata \\
Ne$^{+6}$   &   $2.5_{-1.6}^{+0.2}$~~~~\nodata &   \nodata                                    &  Fe$^{+7}$   &   $1_{-0.8}^{+0.1}$~~~~\nodata   &    \nodata \\
Ne$^{+7}$   &   $6_{-2}^{+0.8}$~~~~\nodata &   \nodata                                      &   Fe$^{+8}$   &   $0.7_{-0.2}^{+0.2}$~~~~\nodata &    \nodata \\
Ne$^{+8}$   &   $10_{-5}^{+1}$~~~~\nodata &   3.5~~~~1.2                                    &   Fe$^{+9}$   &   $1.2_{-0.1}^{+0.6}$~~~~\nodata &   \nodata \\
Ne$^{+9}$   &   $20_{-7}^{+2}$~~~~\nodata   &   10~~~~6.6                                  &    Fe$^{+10}$  &   $1.5_{-0.6}^{+0.1}$~~~~\nodata &   \nodata \\
\cline{1-3}
Mg$^{+4}$   &   $0.5_{-0.5}^{+0.3}$~~~~\nodata &   \nodata                                  &   Fe$^{+11}$  &   $1.2_{-1.1}^{+0.1}$~~~~\nodata  &   \nodata\\
Mg$^{+5}$   &   $0.5_{-0.5}^{+0.4}$~~~~\nodata&  \nodata                                    &   Fe$^{+12}$  &   $0.5_{-0.4}^{+0.1}$~~~~\nodata &   \nodata \\
Mg$^{+6}$   &   $2.0_{-1.6}^{+0.2}$~~~~\nodata &   \nodata                                  & Fe$^{+13}$  &   $0.2_{-0.1}^{+0.1}$~~~~\nodata &    \nodata  \\
Mg$^{+7}$   &   $1.5_{-0.7}^{+0.3}$~~~~\nodata &   \nodata                                  & Fe$^{+14}$  &   $0.2_{-0.2}^{+0.05}$~~~~\nodata   &   \nodata \\
Mg$^{+8}$   &   $2.0_{-1.8}^{+0.2}$~~~~\nodata &   \nodata                                 &  Fe$^{+15}$  &   $0.5_{-0.3}^{+0.1}$~~~~\nodata   &   \nodata\\
Mg$^{+9}$   &   $2.0_{-1.6}^{+0.2}$~~~~\nodata &   \nodata                                  &    Fe$^{+16}$  &   $3_{-1.9}^{+0.3}$~~~~\nodata &     $2.6$~~~~$0.5$\\
Mg$^{+10}$  &   $8.0_{-2.7}^{+0.8}$~~~~\nodata &  $1.3$~~~~$1.4$                            &  Fe$^{+17}$  &   $4.5_{-2.1}^{+0.5}$~~~~\nodata   &    $9.3$~~~~$2.4$   \\
Mg$^{+11}$  &$4.0_{-1.4}^{+0.4}$~~~~$1.2_{-0.6}^{+0.2}$&$\leq5.5$~\tablenotemark{a}~~~~$5.4$ & Fe$^{+18}$  &   $4_{-1.4}^{+0.4}$~~~~\nodata  &     $10.5$~~~~$2.4$    \\
\cline{1-3}
Si$^{+5}$   &   $0.5_{-0.5}^{+1.0}$~~~~\nodata   &   \nodata                                &  Fe$^{+19}$  &   $1.5_{-0.7}^{+0.1}$~~~~\nodata  &  $3.1$~~~~$6.5$ \\
Si$^{+6}$   &    $2.5_{-0.9}^{+1.0}$~~~~\nodata  &   \nodata                                &  Fe$^{+20}$  &   $1_{-0.8}^{+0.1}$~~~~\nodata  & $5.4$~~~~$8.9$  \\
Si$^{+7}$   &   $2.5_{-0.5}^{+2.0}$~~~~\nodata    & \nodata                                 &  Fe$^{+21}$  &   $1_{-0.9}^{+0.1}$~~~~\nodata &   $4.6$~~~~$4.9$ \\
Si$^{+8}$   &   $3.5_{-1.0}^{+0.5}$~~~~\nodata    &   \nodata                               &   Fe$^{+22}$  &   $1_{-0.9}^{+0.1}$~~~~\nodata  &  $0.1$~~~~$10.1$  \\
Si$^{+9}$   &  $4.0_{-0.4}^{+0.6}$~~~~\nodata &   \nodata                                   &    Fe$^{+23}$  &   $0.5_{-0.4}^{+0.4}$~~~~$3_{-3}^{+0.5}$ &   $2.6$~~~~$\leq4.1$~\tablenotemark{a}  \\
Si$^{+10}$  &   $2.0_{-1.4}^{+0.2}$~~~~\nodata &   \nodata                                  &   Fe$^{+24}~\tablenotemark{f}$ &     \nodata~~~~$60_{-26}^{+13}$ &   \nodata \\
& & & Fe$^{+25}~\tablenotemark{f}$ &   \nodata~~~~$120_{-28}^{+52}$ & \nodata

\footnotesize \tablenotetext{a}{~90\% upper limit.}
\tablenotetext{b}{~\citet{lee01} quote $N_{\mathrm{O}^{+5}}\approx3\times10^{17}$ cm$^{-2}$ in the slow component.
 }
 \tablenotetext{c}{~\citet{lee01} quote $N_{\mathrm{O}^{+6}}\geq7\times10^{17}$
cm$^{-2}$ from absorption lines and N$_{\mathrm{O}^{+6}}\approx2.5\times10^{18}$~cm$^{-2}$ from the
edge drop. \citet{turner04} quote $N_{\mathrm{O}^{+6}}\approx2 - 8\times10^{18}$ cm$^{-2}$. Both
refer to the slow component. }
 \tablenotetext{d}{~\citet{lee01} quote $N_{\mathrm{O}^{+7}}\approx 10^{18}$ cm$^{-2}$ from the edge drop. \citet{turner04} quote $N_{\mathrm{O}^{+7}}\approx2.5 - 3\times10^{18}$ cm$^{-2}$. Both
refer to the slow component. }
 \tablenotetext{e}{~\citet{lee01} quote $N_{\mathrm{Fe}^{+0}}\approx4\times10^{17}$ cm$^{-2}$ in the slow component from edge drop.}
  \tablenotetext{f}{~\citet{young05} quote $N_{\mathrm{Fe}^{+24}}=3\times10^{17}$ cm$^{-2}$ and $N_{\mathrm{Fe}^{+25}}=6\times10^{17}$ cm$^{-2}$.}
\enddata
\end{deluxetable}

\begin{deluxetable}{lcccccc}
\tablecolumns{6} \tablewidth{0pt}
\tablecaption{Physical Parameters for The Slow Component: Comparison.
 \label{table5}} \tablehead{
   \colhead{Reference} &
   \colhead{Column Density  } &

   \colhead{Outflow Velocity } &
   \colhead{$v_\mathrm{turb} (b)$} &
   \colhead{Ionization Parameter   } \\

   \colhead{} &
   \colhead{(10$^{21}$~cm$^{-2}$) } &

   \colhead{(\kms)} &
   \colhead{(\kms)} &
   \colhead{log$\xi$ (\cgs) }
}
  \startdata
\multicolumn{1}{l}{\multirow{2}{*}{\citet{otani96}  }}   & 4.6   & \nodata & \nodata  & 1.2\\
                    & 13    & \nodata & \nodata & 1.9\\[0.2 cm]
\multicolumn{1}{l}{\multirow{2}{*}{\citet{reynolds97} }}  & 5     & \nodata & \nodata & 1.3\\
           & 13    & \nodata & \nodata & 1.9\\[0.2 cm]
\multicolumn{1}{l}{\multirow{2}{*}{\citet{lee01} }}     & 5      & \nodata & 100  & 1.2 \\
                   & 30    & \nodata & 100  & 1.9 \\[0.2 cm]
 \citet{sako03}    & 2 \tablenotemark{a}  & --150 $\pm$ 130 & 130  & 0.5 -- 2\\[0.2 cm]
 \citet{turner04}    & \nodata   & 80 $\pm$ 260 & 50 -- 150 & \nodata\\[0.2 cm]
\multicolumn{1}{l}{\multirow{2}{*}{\citet{mckernan07} }}    & 3 $\pm$ 1   & $<$30 & 170  & 0.2 $\pm$ 0.1\\
                      & 4 $\pm$ 1    & $<$15 & 170 & 2.1 $\pm$ 0.1\\[0.2 cm]

\multicolumn{1}{l}{\multirow{2}{*}{\citet{miller08}  }}   & 0.22 $\pm$ 0.01   & \nodata & \nodata & --0.04 $\pm$ 0.16\\
                       & 1.1 $\pm$ 0.8    & \nodata & \nodata & 2.33 $\pm$ 0.05\\[0.2 cm]

 \citet{chelouche08}    & \nodata   & $\sim$ --150~\tablenotemark{b} & \nodata & 0.8 -- 3.1\\[0.2 cm]

 \multicolumn{1}{l}{\multirow{2}{*}{Present Work }}   & 2.3 $\pm$ 0.3   & --100 $\pm$ 50 & 100  & --1.5 -- 0.5\\
                       & 3.0 $\pm$ 0.4    & --100 $\pm$ 50 & 100 & 1.5 -- 3.5
\enddata

\footnotesize \tablenotetext{a}{ \citet{sako03} used twice solar
Fe abundance in order to derive equivalent hydrogen column.}
\tablenotetext{b}{ See Fig.~16 in \citet{chelouche08}.}

\end{deluxetable}

\begin{deluxetable}{lcccccc}

\tablecolumns{6} \tablewidth{0pt}
\tablecaption{Physical parameters of the fast component: Comparison}
 \tablehead{
   \colhead{Reference} &
   \colhead{Column Density  } &

   \colhead{Outflow Velocity } &
   \colhead{$v_\mathrm{turb} (b)$} &
   \colhead{Ionization Parameter   } \\

   \colhead{} &
   \colhead{(10$^{21}$~cm$^{-2}$) } &

   \colhead{(\kms)} &
   \colhead{(\kms)} &
   \colhead{log$\xi$ (\cgs) }
}
  \startdata
 \citet{sako03} & 2~\tablenotemark{a}     & --1900~$\pm$ 140 & 460 & 2 -- 3\\
 \citet{turner04}    & \nodata  & --1970~$\pm$ 160 & 50 -- 150 & \nodata\\
 \citet{young05}     & 20 -- 150~\tablenotemark{b}     & --2000$^{+700}_{-900}$ & 100 -- 500~\tablenotemark{b}  &3.6$^{+0.1}_{-0.2}$ \\
 \citet{mckernan07}    & 30$^{+60}_{-20}$  & --1550$^{+80}_{-150}$ & 170  & 3.7$^{+0.1}_{-0.3}$ \\
 \citet{miller08}    & 21   & --1800 & 500 & 3.85\\
 \citet{chelouche08}    & \nodata   & --2000 & \nodata & 0.3 -- 3.8\\
 Present Work    & 81~$\pm$ 7   & --1900~$\pm$ 150 & 500  & 3.82~$\pm$ 0.03
\enddata

\footnotesize \tablenotetext{a}{ \citet{sako03} used Fe abundance
twice solar to derive equivalent hydrogen column.}
\tablenotetext{b}{ \citet{young05} derive two possible columns
depending on  $v_\mathrm{turb}$. They quote $N_H~\sim 2 - 4
\times\ 10^{22}$~\cmsq\ for $v_\mathrm{turb}~= 500$~\kms\ and $N_H
\sim 1.5\times 10 ^{23}$~\cmsq\ for $v_\mathrm{turb}~= 100$~\kms.}

\label{table6}
\end{deluxetable}

\begin{deluxetable}{lcccc}
 \tablecolumns{6} \tablewidth{0pt}
\tablecaption{Oxygen ionic column densities in the local ($z = 0$)
absorbing component. Respective columns in the slow outflow
intrinsic to \mcg\ are listed for comparison. See
Fig.~\ref{figure4} for the relevant spectrum.} \tablehead{
   \colhead{Charge} &
   \colhead{$N_\mathrm{ion}$ local}&
\multicolumn{2}{c}{Leading K$\alpha$ line/blend} &
   \colhead{$N_\mathrm{ion}$ intrinsic}
\\
   \colhead{State} &
  \colhead{} &
   \colhead{$\lambda _\mathrm{Rest}$}&
  \colhead{$\lambda _\mathrm{Observed}$~\tablenotemark{a}}&
  \colhead{}
\\
  \colhead{} &
    \colhead{(10$^{16}$~cm$^{-2}$)}&
    \colhead{(\AA)}  &
    \colhead{ (\AA)}&
    \colhead{(10$^{16}$~cm$^{-2}$)}
}
\startdata

neutral O~\tablenotemark{b} & $4_{-4}^{+8}$ & 23.523   &23.509 & $10_{-5}^{+5}$    \\
O$^{+1} $ & $10_{-2}^{+20}$  & 23.347~\tablenotemark{c}  & 23.350  &  $8_{-1}^{+16}$    \\
O$^{+2}$ & $1_{-0.7}^{+2}$ & 23.071~\tablenotemark{c}  & 23.065 & $3_{-0.5}^{+6}$ \\
O$^{+3}$  & $1_{-1}^{+2}$ & 22.741~\tablenotemark{c}   & 22.739 &   $6_{-0.6}^{+12}$      \\
O$^{+4}$ & $2_{-1}^{+2}$  & 22.374~\tablenotemark{c}   & 22.345  & $12_{-5}^{+3}$   \\
O$^{+5}$  & $4_{-1}^{+2}$ & 22.019~\tablenotemark{c}   &  22.003 & $12_{-6}^{+1.5}$  \\
O$^{+6}$ & $4_{-1}^{+2}$ &  21.602 &  21.605 &  $60_{-6}^{+7.5}$      \\
O$^{+7}$  & $<$1 & 18.969   &  18.973  & $120_{-24}^{+12}$ \\
\enddata

\footnotesize \tablenotetext{a}{~Uncertainty of $\pm$10~m\AA .}
\tablenotetext{b}{~Strong overlap with intrinsic O$^{+1}$, see
Fig.~\ref{figure4}.} \tablenotetext{c}{~Adjusted to fit data. See
\S \ref{corrections} and Table~\ref{table8}.} \label{table7}
\end{deluxetable}

\begin{deluxetable}{llrccccc}
\tabletypesize{ \scriptsize }
 \tablecolumns{6} \tablewidth{0pt}
\tablecaption{Rest frame wavelengths of oxygen absorption lines and blends determined from \hetgs\ spectrum}
\tablehead{
   \colhead{Ion; Lines} &
\multicolumn{2}{c}{Transition~\tablenotemark{a}} &
  \colhead{$f$-value} &
   \colhead{$\lambda _\mathrm{HULLAC}$} &
  \colhead{$\lambda _\mathrm{HETGS}$~\tablenotemark{b}} &
  \colhead{$\Delta \lambda$} \\
    \colhead{} &
    \colhead{Ground Configuration (J)} &
    \colhead{Upper Configuration (J)} &
    \colhead{} &
    \colhead{(\AA)}  &
    \colhead{(\AA)}  &
    \colhead{ (m\AA)}
}
\startdata
O$^{+1}$ K$\alpha$~\tablenotemark{c} & 1s$^2$2s$^2$2p$_{1/2}$2p$_{3/2}^2$ (J = 3/2)  & 1s2s$^2$2p$_{1/2}^2$2p$_{3/2}^2$ (J = 5/2) & 0.100 & 23.302 & 23.347  & +45  \\
 &   & 1s2s$^2$2p$_{1/2}$2p$_{3/2}^3$ (3/2) & 0.067 & 23.300 & 23.345 &  \\
 &   & 1s2s$^2$2p$_{3/2}^4$ (1/2) & 0.034 & 23.300 & 23.345 &  \\[0.2 cm]

O$^{+2}$ K$\alpha$~\tablenotemark{c} & 1s$^2$2s$^2$2p$_{1/2}^2$ (0) & 1s2s$^2$[2p$_{1/2}^2$2p$_{3/2}$ + 2p$_{3/2}^3$] (1) & 0.125 & 23.108 & 23.071~\tablenotemark{e}  & --37  \\
 &   &  1s2s$^2$[2p$_{1/2}$2p$_{3/2}^2$ + 2p$_{3/2}^3$] (1) & 0.104 & 23.065 &  23.028 &  \\
 &  & 1s2s$^2$[2p$_{1/2}$2p$_{3/2}^2$ + 2p$_{1/2}^2$2p$_{3/2}$] (1) & 0.069 & 22.977 &  22.940 &   \\[0.2 cm]

O$^{+3}$ K$\alpha$~\tablenotemark{c} & 1s$^2$2s$^2$2p$_{1/2}$ (1/2) & 1s2s$^2[$2p$_{1/2}$2p$_{3/2}$ + 2p$_{1/2}^2$] (1/2) & 0.167  & 22.749 &  22.741~\tablenotemark{e}& --8  \\
&  & 1s2s$^2[$2p$_{1/2}$2p$_{3/2}$ + 2p$_{3/2}^2$] (3/2) & 0.077 & 22.747 &  & --6 \\
O$^{+3}$ K$\alpha$&   & 1s2s$^2[$2p$_{1/2}$2p$_{3/2}$ + 2p$_{3/2}^2$] (3/2) & 0.142 & 22.777 &  22.770 & --7  \\[0.2 cm]

O$^{+4}$ K$\alpha$ & 1s$^2$2s$^2$ (0) & 1s2s$^2$2p$_{3/2}$ (1)& 0.539 & 22.337 & 22.374~\tablenotemark{e}  & +37 \\
O$^{+4}$ K$\beta$ &  & 1s2s$^2$3p$_{3/2}$ (1)& 0.112 & 19.924& 19.968  & +44 \\
O$^{+4}$ K$\gamma$ & & 1s2s$^2$4p$_{3/2}$ (1)& 0.046 & 19.324 & 19.357  & +33 \\[0.2 cm]

O$^{+5}$ K$\alpha$~\tablenotemark{c} &1s$^2$2s (1/2) & 1s2s2p$_{3/2,1/2}$ (3/2, 1/2)& 0.349, 0.173 & 22.013 & 22.019~\tablenotemark{d} & +6 \\
O$^{+5}$ K$\beta$~\tablenotemark{c}& & 1s2s3p$_{3/2}$ (3/2, 1/2)& 0.064, 0.032 & 19.341 & 19.379 & +38 \\
O$^{+5}$ K$\beta$~\tablenotemark{c}& & 1s2s3p$_{3/2,1/2}$ (3/2, 1/2)& 0.025, 0.012 & 19.136 & 19.180  & +44  \\
O$^{+5}$ K$\gamma$~\tablenotemark{c}& & 1s2s4p$_{3/2}$ (3/2, 1/2)& 0.027, 0,014 & 18.606 & 18.587   & --19 \\
O$^{+5}$ K$\delta$~\tablenotemark{c}&  & 1s2s5p$_{3/2}$ (3/2, 1/2) & 0.015, 0.007 &18.290 & 18.270   & --20
\enddata

\tablenotetext{a}{~Square brackets indicate significant
configuration mixing.} \tablenotetext{b}{~With an accuracy of $\pm
10$~m\AA.} \tablenotetext{c}{~Unresolved blend shifted uniformly.}
\tablenotetext{d}{~\citet{schmidt04} give 22.019 $\pm$ 0.003 \AA\
for this line.} \tablenotetext{e}{~\citet{gu05} report blends at
22.374 $\pm$ 0.003 \AA, 22.741 $\pm$ 0.004~\AA, and 23.071 $\pm$
0.004~\AA\ that include, respectively, the relevant O$^{+4}$,
O$^{+3}$, and O$^{+2}$ K$\alpha$ lines.}
 \label{table8}
\end{deluxetable}

\end{document}